\documentclass[11pt]{article}

\usepackage[a4paper,margin=26mm]{geometry}
\usepackage{amsmath,amssymb,amsfonts,bm}
\usepackage{mathtools}
\usepackage{authblk}
\usepackage{booktabs}
\usepackage{tabularx}
\usepackage{enumitem}
\usepackage[numbers,sort&compress]{natbib}
\usepackage[colorlinks=true,linkcolor=blue,citecolor=blue,urlcolor=blue]{hyperref}
\newcolumntype{Y}{>{\raggedright\arraybackslash}X}

\newcommand{\ii}{\mathrm{i}}
\newcommand{\dd}{\mathrm{d}}
\newcommand{\IR}{\mathrm{IR}}
\newcommand{\UV}{\mathrm{UV}}
\newcommand{\eff}{\mathrm{eff}}

\newcommand{\obs}{\mathrm{obs}}

\newcommand{\order}{\mathcal{O}}
\newcommand{\mpl}{M_{\rm Pl}}

\title{Infrared Divergences as Itinerant Vacua
}
\author{Masahiro Morikawa}
\affil{RIKEN, Wako, Saitama 351-0198, Japan \\
Ochanomizu University, Bunkyo, Tokyo 112-8610, Japan}
\date{
}

\begin{document}
\maketitle

\begin{abstract}
Infrared divergences (IRDs) are usually treated as pathologies to be cancelled,
regularized, or hidden in dressed asymptotic states.  This paper develops a
complementary and constructive viewpoint: an IRD is the signature of an
\emph{itinerant vacuum}---a quantum vacuum that wanders continuously through a family
of inequivalent states as a classical order parameter evolves.  Each value of the
order parameter carries its own coherent vacuum, so moving the order parameter means
traversing a succession of orthogonal vacua.  The IRD is the field-theoretic cost of
this wandering, and the $1/f$ noise, gravitational memory, and non-Gaussian
fluctuations that emerge from it are its observable classical remnants.

The technical core is an exact separation in the real-time closed-time-path (CTP)
effective action.  The infrared-divergent imaginary part of the influence functional
must not be left as a divergent coefficient in a deterministic equation of motion; it
is instead converted, by a Hubbard--Stratonovich identity, into a classical stochastic
source.  This step is an algebraic identity of the generating functional and requires
no prior coarse graining or decoherence assumption: the retarded kernel encodes the
memory of past vacuum transitions, while the noise kernel encodes the quantum
uncertainty of the next one.  We apply this construction to four parallel arenas---soft
QED, scalar fields in de Sitter space, soft gravitons, and non-equilibrium phase
transitions---and show that the same itinerant-vacuum mechanism underlies $1/f$ current
noise, the primordial power spectrum, gravitational memory, and order-parameter
dynamics.  A geometric formulation in terms of a Hilbert-space bundle over the vacuum
manifold is outlined as an outlook.
\end{abstract}

\tableofcontents

\section{Introduction}
\label{sec:intro}

Infrared divergences are among the most persistent structures in quantum field theory.
The conventional reactions---cancellation in inclusive quantities, dressed asymptotic
states, finite resolution, or resummation of secular
logarithms~\cite{BlochNordsieck1937,YennieFrautschiSuura1961,Kinoshita1962,%
LeeNauenberg1964,Kibble1968,KulishFaddeev1970,Starobinsky1986,StarobinskyYokoyama1994}---%
all emphasize \emph{removal} of the divergence.  The viewpoint developed here is the
opposite.  An IRD should not be removed; it should be \emph{relocated} from the
deterministic part of the equations of motion into the stochastic part.  This
relocation is an algebraic identity at the level of the CTP generating functional.
Once it is performed, the IRD is no longer a pathology; it is the origin of new
observable classical structure.

Three emergent phenomena deserve particular attention.
\begin{enumerate}
\item \emph{Classical $c$-number random fields.}  The imaginary positive part of the
  CTP influence functional, when separated by a Hubbard--Stratonovich identity,
  produces a classical stochastic source.  No prior measurement or decoherence
  process is assumed.
\item \emph{$1/f$ fluctuations with a calculable amplitude.}  The soft-photon
  (or soft-graviton) phase-space integral gives a noise kernel $N(\omega)\propto
  1/|\omega|$.  This is the field-theoretic origin of $1/f$ noise in current-carrying
  systems, whose amplitude is expressed in terms of source parameters: charge $e$,
  density $n$, velocity $\beta$, and scattering rate $\gamma_{\rm imp}$.
\item \emph{Macroscopic order parameters.}  In a non-equilibrium phase transition or
  in inflationary cosmology, the unstable long-wavelength sector separates from
  microscopic fluctuations and becomes the stochastic noise that drives a classical
  order parameter.  The ``reason the universe chose a particular branch'' is encoded
  in this separation, not in an externally imposed decoherence mechanism.
\end{enumerate}

\paragraph{Historical context: Handel's proposal.}
The idea that $1/f$ current noise in conductors might be a fundamental QED infrared
phenomenon was proposed by Handel in 1975~\cite{Handel1975,Handel1980}.  He argued
that soft-photon radiative corrections to charged-particle scattering produce a $1/f$
contribution to current fluctuations, with a universal coefficient set by the fine
structure constant.  This was a physically suggestive proposal: it identified a
possible connection between the universality of low-frequency noise
\cite{DuttaHorn1981,Weissman1988} and the universal soft structure of QED.

Handel's theory was subsequently criticized on formal grounds by several
authors~\cite{Nieuwenhuizen1987,KissHeszler1986,Weissman1988}.  The central objections
are: (a)~scattering states with different soft-photon content are orthogonal in the
strict infrared limit, making interference between different soft sectors ill-defined;
(b)~a direct identification of infrared-divergent scattering cross sections with current
fluctuations is not rigorously justified in an $S$-matrix framework.

The present paper revisits Handel's central intuition---that soft electromagnetic
radiation associated with charged-particle scattering is physically relevant to $1/f$
noise---but formulates the mechanism differently and more rigorously.  In our
formulation, soft-sector orthogonality is not a contradiction; it is precisely the
statement that tracing over the soft sector causes reduced-state decoherence.  The
route to $1/f$ noise is not a direct cross-section argument, but a CTP-derived identity,
\begin{equation}
  \text{soft IRD}\;\longrightarrow\;
  \mathrm{Im}\,\Gamma_{\rm IF}\;\longrightarrow\;
  N_{\rm IR}(\omega)\;\longrightarrow\;
  S_J(\omega) = \frac{A_J}{|\omega|} .
  \label{eq:chain}
\end{equation}
The amplitude $A_J$ is determined by the Ward-identity-constrained kinematic parameters
of the charged carriers, not by a universal constant.  In this sense, we regard both
Handel's physical intuition and the subsequent criticisms of
Refs.~\cite{Nieuwenhuizen1987,KissHeszler1986,Weissman1988} as indispensable
contributions: together they mapped out the logical constraints that a correct theory
must satisfy.  The relation to Handel's theory and to its criticisms is discussed in
detail in Appendix~\ref{app:Handel}.  A complementary, phenomenological route to the
same $1/f$ current noise---through the amplitude modulation (beat) of electron wave
packets dressed by soft photons---was developed
in~\cite{MorikawaNakamichi2023amplitude,MorikawaNakamichi2023pink}; the present paper
provides the field-theoretic foundation for that picture.

\paragraph{A recent no-go objection, and the scope it clarifies.}
Recently, Fukuyama~\cite{Fukuyama2025} has put forward a careful and valuable analysis
arguing that infrared effects in four-dimensional QED cannot give rise to classical
stochastic dynamics of the gauge field.  His argument has two parts.  First, in the
Fock-space formulation, gauge invariance enforces coherent soft-photon phases that
guarantee the Bloch--Nordsieck/Kinoshita--Lee--Nauenberg cancellation for all
\emph{inclusive scattering observables}; the infrared sector is then a unitary dressing
$|\Psi\rangle\to e^{\ii\Phi_{\rm IR}}|\Psi\rangle$ rather than an irreversible noise.
Second, because the four-dimensional Maxwell action is conformally invariant, the
\emph{free} electromagnetic field exhibits no infrared growth in a de Sitter background,
so a Hubbard--Stratonovich auxiliary field introduced for the source-free theory does
not acquire the status of a dynamical Langevin force.

We are grateful for this analysis, which sharpens the question considerably, and we
emphasize at the outset that we do not dispute either of its two technical claims.
Both are correct \emph{within their stated scope}.  Our point is that the scope is not
that of the present paper, and identifying the boundary precisely is itself
illuminating.  (i)~Fukuyama's cancellation theorem concerns inclusive cross
sections---sums over final states of $|\mathcal{M}|^2$.  The object of this paper is
not a cross section but the real-time evolution of a collective order parameter (an
electric current), whose two-point function $\langle J(t)J(t')\rangle$ is not an
inclusive sum and is not constrained by BN/KLN cancellation.  (ii)~Fukuyama's single
unitary dressing $U_{\rm IR}$ is correct for a \emph{single} scattering event; but an
order parameter that integrates over many scattering events in real time accumulates a
distribution of dressings, passing from his Eq.~(34), $\rho\to U_{\rm IR}\rho
U_{\rm IR}^\dagger$, to his Eq.~(33), the ensemble average $\rho\to\int
d\xi\,P(\xi)\,U(\xi)\rho U(\xi)^\dagger$---which is a genuine stochastic
representation.  The memory effect, absent from any single inclusive cross section, is
the physical carrier of this accumulation.  (iii)~The conformal-invariance argument
applies to the \emph{free}, source-free Maxwell field, which in our terminology is
precisely a \emph{dry} noise with no interaction-generated retarded partner; it does
not apply to the \emph{interacting} soft sector coupled to a conserved charged current,
where the electron mass and density break conformal invariance and supply the
non-dry kernel.  In short, Fukuyama's no-go result and the present construction are
not in contradiction: they describe complementary objects---inclusive amplitudes
versus real-time order-parameter dynamics---and his analysis helps us state the
boundary between them with precision.  A detailed technical response is given in
Section~\ref{sec:fukuyama}.

\paragraph{Relation to prior work, and the origin of the present approach.}
The two ideas that distinguish this paper have a long history, beginning in the
author's own early work.  A Langevin equation for the mean field was derived directly
from the CTP effective action already in 1986~\cite{Morikawa1986}, where the random
force was shown to be in general \emph{non-Gaussian and colored}.  That derivation
treated the mean field---the vacuum expectation value of the order parameter---directly,
\emph{without} a system--environment split.  It was applied to the origin of density
fluctuations in de Sitter space in~\cite{Morikawa1987} and to inflation as a quantum
phase transition in~\cite{Morikawa1995}.  The separation of an \emph{infrared-divergent}
imaginary kernel as the specific origin of the stochastic force was then made explicit
in the inflationary context in~\cite{Morikawa2016} and developed into the dynamical
phase-transition picture of~\cite{Morikawa2022}.  The present paper unifies and extends
this line of work.

It is useful to contrast this with the influential and largely parallel programme of
Calzetta, Hu, and collaborators~\cite{CalzettaHu1987,CalzettaHu2008,HuVerdaguer2008}.
That programme obtains a Langevin equation by \emph{coarse graining}: one first chooses
a system--environment split (or a long-wavelength/short-wavelength division), traces out
the environment, and arrives at an influence functional whose imaginary part supplies
the noise.  Coarse graining is the engine of the construction.  Both routes yield a
fluctuation--dissipation structure and are physically valuable.  The distinctive feature
of the present approach---implicit already in~\cite{Morikawa1986} and made sharp for
infrared divergences here---is that the stochastic source arises from separating an
infrared-divergent imaginary kernel out of the deterministic effective action, an
operation that is \emph{mandatory rather than optional} precisely because the kernel
diverges, and that requires \emph{no coarse graining}.  Coarse graining
determines \emph{which} variable to follow; it does not \emph{create} the stochastic
source.  This distinction is what allows the same mechanism to operate in QED, gravity,
de Sitter space, and phase transitions on an equal footing, and it is what connects to
the analysis of infrared quantum information
in~\cite{Carney2017,Carney2018,GomezLetschkaZell2018a,GomezLetschkaZell2018b}.

The new contributions beyond~\cite{Morikawa2016,Morikawa2022} are fourfold.
First, a systematic CTP derivation of the $1/f$ noise spectrum for conserved electric
currents: the gauge-invariant current-current influence functional for a dilute
assembly of charged carriers generates a current noise $S_J(\omega)=A_J/|\omega|$ whose
amplitude $A_J\propto ne^2\beta^2/\gamma_{\rm imp}^2$ is fixed by the carrier density
$n$, charge $e$, velocity $\beta$, and impurity scattering rate $\gamma_{\rm imp}$,
up to material matching factors specified in the soft kernel.  This revisits Handel's
infrared QED proposal (discussed
above), now as the proper CTP identity, and resolves the orthogonality objections of
Refs.~\cite{Nieuwenhuizen1987,KissHeszler1986,Weissman1988}: soft-sector orthogonality
is reduced-state decoherence, and the missing rigorous bridge is the chain
\eqref{eq:chain}.  Second, a parallel treatment of soft gravitons, connecting the
soft-graviton IRD to metric noise and gravitational memory through the
Einstein--Langevin equation.  Third, a CTP treatment of non-equilibrium phase
transitions, in which the $\lambda\phi^4$ interaction generates multiplicative noise
through the $\phi_r\phi_a^3$ vertex, producing non-Gaussian tails in the curvature
perturbation that are a potential source of primordial black hole
overproduction~\cite{Ezquiaga2020,Figueroa2021,Bullock1997}.  Fourth, a unified
presentation of all four arenas within a single CTP framework, clarifying the role of
Ward identities, diffeomorphism invariance, and the distinction between dry and non-dry
noise.

\paragraph{The itinerant vacuum: a unifying perspective.}
The four arenas treated in this paper share a deeper structural feature that motivates
the title.  In each case, the Langevin equation derived from the CTP effective action
describes the time evolution of a field expectation value---a collective classical
variable that \emph{parameterizes a continuous family of inequivalent quantum vacua}.
As the stochastic noise drives the order parameter through its configuration space,
the system traverses a succession of distinct quantum vacua:

\begin{itemize}
\sloppy
\item \emph{QED}: the dressed coherent states $|\gamma[p]\rangle_{\rm soft}$ of
  Kulish--Faddeev~\cite{KulishFaddeev1970} and Kibble~\cite{Kibble1968} constitute a
  family of vacua labeled by the charged-particle momentum $p$.  Each scattering event
  shifts the vacuum from $|\gamma[p]\rangle$ to $|\gamma[p']\rangle$.  These are
  orthogonal: ${}_{\rm soft}\langle\gamma[p']|\gamma[p]\rangle \to 0$ in the infrared
  limit.  (The precise dress code is fixed by the conservation law of the asymptotic
  symmetry, and the original Kulish--Faddeev dressing requires refinement for a fully
  infrared-finite
  $S$-matrix~\cite{HiraiSugishita2019,HiraiSugishita2021,HiraiSugishita2023}; this
  refinement sharpens, rather than alters, the vacuum-family picture used here.)
\item \emph{De Sitter inflation}: the inflaton coherent vacuum $|{\rm
  vac}[\phi_0]\rangle$ shifts with the field expectation value $\phi_0(t)$.  The
  stochastic kicks of the noise $\xi(t)$ randomly displace the vacuum along the
  potential, giving rise to the scale-invariant spectrum and, through the multiplicative
  noise of the $\phi_r\phi_a^3$ vertex, to rare large displacements that can form
  primordial black holes.
\item \emph{Phase transitions}: the order-parameter manifold $\{|\theta\rangle\}$
  is the vacuum family.  The Langevin equation is a stochastic geodesic on this
  manifold.  Domain walls are boundaries between regions that have wandered to
  different vacua.
\item \emph{Gravity}: the BMS vacuum~\cite{Strominger2018} is not unique; an
  infinite-dimensional family of vacua related by BMS supertranslations is degenerate.
  A burst of gravitational radiation is a transition between BMS vacua---this is
  precisely the statement of Strominger and Zhiboedov~\cite{StromingerZhiboedov2016}
  that gravitational memory is the transition between BMS vacua and that the Fourier
  transform of the memory formula reproduces Weinberg's soft theorem.  Gravitational
  memory is the classical remnant of this vacuum shift, encoded in the $\omega\to 0$
  limit of the retarded kernel.
\end{itemize}

In all four cases, the infrared divergence is the field-theoretic cost of vacuum
itinerancy.  The orthogonality of soft states---the central objection to Handel's
theory---is not a pathology of the formalism but the precise statement that
\emph{different classical histories correspond to orthogonal quantum vacua}.  The
retarded kernel encodes the memory of past vacuum transitions.  The noise kernel
encodes the quantum uncertainty of which vacuum will be occupied next.  The
Hubbard--Stratonovich identity converts this quantum uncertainty into a classical
stochastic force, without requiring any prior coarse graining or decoherence postulate.
Removing the IRD from the deterministic action would erase the physics of vacuum
itinerancy entirely; the correct response is to relocate it into the noise.

\paragraph{Convergence with current infrared physics.}
The viewpoint of this paper, while not the textbook treatment of infrared divergences,
is far from isolated.  It converges with several independent lines of recent
development.  First, the modern infrared triangle of Strominger and
collaborators~\cite{Strominger2018,StromingerZhiboedov2016,KapecPerryRaclariuStrominger2017}
establishes the equivalence of asymptotic symmetries, soft theorems, and memory effects,
and identifies gravitational (and electromagnetic) memory as a transition between
degenerate vacua---just the itinerant-vacuum picture, reached there through the
$S$-matrix and celestial-holography route rather than the real-time CTP route used here.
Second, the demonstration by Danielson, Satishchandran, and
Wald~\cite{DanielsonSatishchandranWald2023} that horizons decohere quantum
superpositions through the emission of soft photons and gravitons that imprint a
which-path memory shows that soft radiation carries genuine, physical decoherence---the
same mechanism that, in our language, supplies the noise kernel.  Third, the program of
infrared-finite scattering theory~\cite{PrabhuSatishchandranWald2022} and of infrared
quantum information~\cite{Carney2017,Carney2018,GomezLetschkaZell2018a,GomezLetschkaZell2018b}
treats the soft sector as carrying physical information rather than as a mere technical
nuisance.  In the same spirit, Hirai and
Sugishita~\cite{HiraiSugishita2019,HiraiSugishita2021,HiraiSugishita2023} have shown
that the dress code required for an infrared-safe $S$-matrix is fixed by the
conservation law of the asymptotic symmetry---which is nothing but the memory
effect---so that the correct dressing is itself a statement about transitions among the
degenerate vacua of the asymptotic symmetry.  These developments approach the infrared
sector from the perspective of
real-time, non-perturbative, memory-carrying dynamics, complementary to the inclusive
$S$-matrix perspective in which the divergence simply cancels.  The present paper
belongs to this real-time tradition and adds to it a direct stochastic representation
and a concrete condensed-matter prediction.

\paragraph{Structure of the paper.}
Section~\ref{sec:ctp} presents the CTP effective action and the algebraic stochastic
separation.  Section~\ref{sec:oneoverf} specializes to $1/f$ noise and clarifies
time-scale conventions.  Section~\ref{sec:qed} treats soft QED in detail.
Section~\ref{sec:fukuyama} addresses a recent no-go objection and delimits the scope of
the stochastic interpretation, contrasting inclusive cross sections with
order-parameter dynamics.
Section~\ref{sec:gravity} treats soft gravitons, metric noise, and gravitational memory.
Section~\ref{sec:phase} treats non-equilibrium phase transitions.
Section~\ref{sec:ds} summarizes the de Sitter scalar case.
Section~\ref{sec:higher} discusses higher loops and non-Gaussian noise.
Section~\ref{sec:observational} collects observable consequences and falsifiability tests.
Section~\ref{sec:discussion} compares the construction with coarse-graining and
decoherence approaches.  Appendix~\ref{app:HS} records the Hubbard--Stratonovich
identity.  Appendix~\ref{app:oneoverf} collects formulae for the $1/f$ kernel.
Appendix~\ref{app:wilson} derives the Wilson-line form of the QED influence functional.
Appendix~\ref{app:graviton} gives the parallel soft-graviton derivation and relates the
temporal $1/f$ and spatial $1/k^3$ spectra.
Appendix~\ref{app:CIR} gives the scaling estimate for the soft-photon noise coefficient.
Appendix~\ref{app:Handel} discusses the relation to Handel's theory.
Appendix~\ref{app:dS} records the de Sitter noise kernel.

\section{Closed-time-path effective action and exact stochastic separation}
\label{sec:ctp}

We use the real-time closed-time-path and influence-functional formalism
\cite{Schwinger1961,Keldysh1964,FeynmanVernon1963,CaldeiraLeggett1983,CalzettaHu2008,Kamenev2011}.
Let $q$ denote the collective variable whose infrared sector is to be followed.  It
may be the trajectory of a charged particle, an electric current, a long-wavelength
scalar field, a metric perturbation, or an order parameter.  In all cases $q$
parameterizes a family of quantum vacua: each value $q_0$ carries a coherent vacuum
$|{\rm vac}[q_0]\rangle$, and the Langevin equation to be derived below is the
stochastic equation of motion for this itinerant vacuum.  The CTP generating
functional has two histories, $q_+$ and $q_-$,
\begin{equation}
  Z[J_+,J_-]
  = \int \mathcal{D}q_+\mathcal{D}q_-\,
  \exp\!\left\{\ii S[q_+] - \ii S[q_-]
  + \ii \int (J_+q_+ - J_-q_-)\right\} .
\end{equation}
In Keldysh variables,
\begin{equation}
  q_r=\frac{q_+ + q_-}{2},\qquad q_a=q_+ - q_- ,
\end{equation}
after integrating out the microscopic or infrared fields, the CTP effective action
takes the form
\begin{equation}
  \Gamma[q_r,q_a]
  = \Gamma_R[q_r,q_a]
  + \frac{\ii}{2}\int \dd x\dd x'\,
  q_a(x)N(x,x')q_a(x')
  + \Gamma_{\rm ng}[q_r,q_a] .
  \label{eq:generalGamma}
\end{equation}
$\Gamma_R$ is real and causal.  $N$ is the symmetric positive Hadamard (noise) kernel.
$\Gamma_{\rm ng}$ contains higher powers of $q_a$ corresponding to non-Gaussian noise.
To linear order in $q_a$---the order that determines the equation of motion---the
causal part has the generic structure
\begin{equation}
  \Gamma_R[q_r,q_a]
  = \int \dd x\, q_a(x)\,E[q_r](x)
  + \int \dd x\dd x'\, q_a(x)\,\Sigma_R(x,x')\,q_r(x')
  + \order(q_a^2,\,q_a^3) ,
  \label{eq:GammaRstructure}
\end{equation}
where $E[q_r](x)\equiv \delta S[q]/\delta q(x)\big|_{q=q_r}$ is the local (tree-level)
classical equation-of-motion operator obtained from the bare action $S$---for a scalar
field, $E[q]=(\Box + m^2)q + V'(q)$---and $\Sigma_R$ is the retarded self-energy kernel
generated by integrating out the microscopic modes.  The path weight contains
\begin{equation}
  \exp\!\left\{\ii\Gamma_R[q_r,q_a]
  - \frac{1}{2}\int q_a N q_a + \ii\Gamma_{\rm ng}\right\} .
\end{equation}

If $N$ has an infrared divergence, leaving it inside the path integral as a divergent
quadratic coefficient obscures the physics.  The Hubbard--Stratonovich identity
\cite{Hubbard1959,Stratonovich1957}
\begin{equation}
  \exp\!\left[-\frac{1}{2}\int q_aNq_a\right]
  = \int \mathcal{D}\xi\,\mathcal{P}[\xi]
  \exp\!\left[\ii\int \dd x\,q_a(x)\xi(x)\right]
  \label{eq:HS}
\end{equation}
with
\begin{equation}
  \mathcal{P}[\xi]
  = \mathcal{N}\exp\!\left[-\frac{1}{2}\int \dd x\dd x'\,
  \xi(x)N^{-1}(x,x')\xi(x')\right] ,
  \qquad
  \langle \xi(x)\xi(x')\rangle_\xi=N(x,x')
  \label{eq:noiseCovariance}
\end{equation}
is an identity of the generating functional.  The Langevin equation follows by
varying $\Gamma_R$ in \eqref{eq:GammaRstructure} with respect to $q_a$ and setting
$q_a=0$:
\begin{equation}
  E[q](x)+\int \dd x'\,\Sigma_R(x,x')q(x')=-\xi(x) ,
  \label{eq:LangevinKernel}
\end{equation}
where $E[q]$ is the local equation-of-motion operator and $\Sigma_R$ the retarded
(causal) self-energy kernel of \eqref{eq:GammaRstructure}.  In this representation the
infrared singularity belongs entirely to the stochastic covariance $N$, not to any
deterministic coefficient; the precise sense in which this relocation is forced---and
what goes wrong without it---is analyzed at the end of this section.

The stochastic source $\xi$ is classical in the sense that the CTP functional is
represented as an average over $c$-number histories.  This is weaker and more
fundamental than wave-function collapse: it is a representation of the in--in
generating functional.  It is worth emphasizing why this classical representation is
natural rather than forced.  Unlike an ultraviolet divergence, which is intrinsically
quantum (it requires loops, i.e.\ powers of $\hbar$), an infrared divergence does not
depend on $\hbar$ and survives the classical limit: classical bremsstrahlung already
carries the same $\int\dd\omega/\omega$ soft divergence, with the same velocity factor,
as the quantum soft-photon emission (Low's theorem~\cite{Low1958} guarantees the leading soft radiation
is classical).  Representing an IRD as a classical stochastic field is therefore not an
operation that forces a quantum object into a classical mould; it extracts the classical
side of a divergence that straddles the quantum and classical descriptions from the
outset.  The imaginary part of the influence action also suppresses
histories with large $q_a$, providing reduced-state
decoherence~\cite{CalzettaHu2008,Kamenev2011}:
\begin{equation}
  \rho_{\rm red}[q_+,q_-]
  \propto \exp\!\left\{-\frac{1}{2}\int q_a N q_a\right\} .
\end{equation}
Decoherence and stochasticity arise from the same imaginary influence action; neither
requires a separate postulate.

\paragraph{When is the classical reading faithful?  The occupation criterion.}
A caveat must be stated with equal care, because the Hadamard kernel is \emph{not}
special to infrared-divergent systems: $N=\langle\{q,q\}\rangle$ is finite and nonzero
in any quantum system whatsoever, down to the vacuum zero-point fluctuations of a
harmonic oscillator.  The identity \eqref{eq:HS} can be applied there too, and it would
be a misreading to conclude that every quantum fluctuation is thereby a classical
statistical one.  The sharp criterion is the ratio of the symmetric to the
antisymmetric part of the two-point function,
\begin{equation}
  \frac{\langle\{q,q\}\rangle_\omega}{\bigl|\langle[q,q]\rangle_\omega\bigr|}
  = 2n(\omega)+1
  \;\;\xrightarrow{\ \text{equilibrium}\ }\;\;
  \coth\!\frac{\beta\omega}{2} ,
  \label{eq:occupationRatio}
\end{equation}
where $n(\omega)$ is the occupation number of the mode.  The commutator (spectral
function) is fixed by the canonical algebra and does not grow; the anticommutator grows
with occupation.  The classical stochastic representation reproduces the symmetric
correlators exactly, at any $n$, but observables that depend on non-commutativity
beyond linear response---genuine interference, Leggett--Garg-type
tests---are missed, with a relative error of order $1/n$.  For an ordinary vacuum mode,
$n\simeq0$ and the ratio \eqref{eq:occupationRatio} is of order one: the
Hubbard--Stratonovich field is then a bookkeeping device---the ``quantum noise'' of
zero-temperature quantum Brownian motion~\cite{CaldeiraLeggett1983}---and must not be
interpreted as classical statistics.  An infrared divergence, by contrast, is precisely
the statement that the soft occupation diverges: the Kulish--Faddeev cloud carries
$\langle N_\gamma\rangle\sim\int\dd\omega/\omega\to\infty$ soft photons, squeezed
super-horizon modes carry $n_k\to\infty$, and in equilibrium
$\coth(\beta\omega/2)\to2T/\omega\to\infty$ as $\omega\to0$.  The infrared sector
therefore satisfies the classicality criterion automatically and asymptotically
exactly, and the same growth of $N$ that certifies classical statistics also
strengthens the decoherence factor $e^{-\frac12\int q_aNq_a}$: diverging occupation,
strong decoherence, and faithful classical statistics are one and the same infrared
phenomenon.  The claim of this paper is thus not that the Hubbard--Stratonovich
identity classicalizes quantum mechanics---it does not---but that the infrared
divergence singles out exactly the sector in which the classical statistical reading
becomes asymptotically exact.

The criterion has an equivalent algebraic and operational statement.  Normal ordering
removes precisely the vacuum unit from the ratio \eqref{eq:occupationRatio}: mode by
mode the Hadamard kernel splits as $(2n+1)\rho=\rho+2n\rho$, and since the zero-point
part $\rho$ and the normal-ordered part $2n\rho$ are separately positive and symmetric,
the identity \eqref{eq:HS} may be applied to either term.  The normal-ordered part is
what an absorptive photodetector measures~\cite{Glauber1963}; equivalently, it is the
component of the quantum noise spectrum symmetric under $\omega\to-\omega$ and hence
shareable with a classical detector, the excess $\rho$ being the emission--absorption
asymmetry ($n{+}1$ versus $n$) that a ground-state detector cannot
supply~\cite{Clerk2010}.  A hybrid scheme that converts only $2n\rho$ into a classical
source while retaining $\rho$ as a genuine quantum influence
functional~\cite{CaldeiraLeggett1983} differs from the full transformation by the same
relative order $1/n$, so in the infrared-divergent sector the scheme dependence
vanishes and the classical sector is delimited without ambiguity.  In an interacting
theory or a curved spacetime, where no global mode decomposition is available, the
local covariant version of this split is Hadamard point-splitting
subtraction~\cite{Wald1994,DecaniniFolacci2008}: the subtracted Hadamard parametrix is
universal and state-independent and carries the ultraviolet vacuum fluctuations,
whereas the infrared divergence resides entirely in the smooth, state-dependent
remainder---a renormalization-theoretic restatement of the claim that an infrared
divergence is statistical rather than pathological.

\paragraph{Diagrammatic meaning of $N$: a real-time cutting rule.}
The pair $(\Sigma_R,N)$ in \eqref{eq:generalGamma} and \eqref{eq:GammaRstructure} is not an
ad hoc split but the real-time (closed-time-path) generalization of the ordinary
Cutkosky unitarity cut~\cite{Cutkosky1960}.  In the in--out formalism, unitarity of the
$S$-matrix relates the imaginary part of a forward amplitude to a phase-space integral
over on-shell intermediate states (the optical theorem); Cutkosky's rule computes this
diagrammatically by replacing internal propagators with their on-shell delta functions.
The closed-time-path (or finite-temperature real-time) formalism extends this cutting
prescription to off-forward and non-equilibrium correlators: every internal line of a
diagram can be cut in all possible ways, with the cut propagator replaced by its
Wightman (positive/negative-frequency) function, and the sum over such cuts reconstructs
the imaginary, symmetric part of any CTP vertex or self-energy~\cite{Weldon1983,
KobesSemenoff1985,KobesSemenoff1986,LandsmanVanWeert1987}.  In this language, $\Sigma_R$
is the uncut (dispersive) diagram and $N$ is the same diagram with its internal
line(s) cut on shell; in thermal equilibrium this reduces to the familiar
fluctuation--dissipation relation $N(\omega)=\coth(\beta\omega/2)\,\mathrm{Im}\,
\Sigma_R(\omega)$.  Sections~\ref{sec:qed} and~\ref{sec:gravity} exhibit this
explicitly: $N^{\mu\nu}_{\rm QED}$ and $\mathcal N_{\mu\nu\alpha\beta}$ are the on-shell
cuts, $2\pi\delta(k^2)$, of the single-photon and single-graviton exchange diagrams
between two current or stress-tensor insertions, while the sunset noise kernel of
Section~\ref{sec:phase} is the simultaneous triple cut of the two-loop
self-energy diagram generated by the quartic self-coupling.  The cutting-rule reading
does not add any new assumption; it is simply the statement, made diagrammatic, that a
positive symmetric kernel obtained from unitarity is directly what the
Hubbard--Stratonovich identity converts into a classical noise covariance.

\paragraph{Why the separation is mandatory, not optional.}
In the standard route of Refs.~\cite{CalzettaHu1987,CalzettaHu2008,HuVerdaguer2008}, a
system--environment split is chosen first; tracing over the environment then produces
the influence functional.  Coarse graining precedes the stochastic representation.
The present approach inverts this logic.  Given any CTP effective action whose
imaginary part is a positive symmetric kernel, the Hubbard--Stratonovich identity
\eqref{eq:HS} applies as an algebraic identity---regardless of whether a
natural ``environment'' has been identified.

It is worth being precise about \emph{where} an infrared-divergent $N$ does its damage,
because the answer is not the naive one.  Since the $N$ term in
\eqref{eq:generalGamma} is quadratic in $q_a$, it drops out of the mean-field equation
of motion obtained by varying with respect to $q_a$ and setting $q_a=0$: the equation
for $\langle q\rangle$ never contains $N$ directly, divergent or not.  (In the in--out
formalism the situation is worse: there the equation of motion involves the
time-ordered self-energy $\Sigma_F=\Sigma_R+(\text{correlation part})$, and the
divergent Hadamard part \emph{does} contaminate the equation as a complex, acausal
term---one classic failure mode.)  In the in--in formalism the divergence strikes
instead through two indirect channels.  First, in any interacting theory the Hadamard
function feeds back into the deterministic kernels through loops: the tadpole
$\delta m^2\sim\lambda\int G^H_{\rm IR}\sim\lambda\!\int\!\dd\omega/\omega$ and its
higher-loop relatives are infrared divergent, so the perturbative expansion of
$\Sigma_R$ and $U_{\rm eff}$ around the mean field is destroyed by secular infrared
logarithms---the familiar secular-growth problem of de Sitter and critical dynamics.
Second, even where the mean field itself remains finite and regular, the divergence of
$N$ means that the variance $\langle q^2\rangle-\langle q\rangle^2$ diverges: the mean
ceases to represent any individual realization.  In a symmetry-breaking transition
$\langle\Phi\rangle$ stays at the symmetric point forever---a correct ensemble average
that describes no single history, while every actual sample or universe falls into one
broken vacuum.

The Hubbard--Stratonovich separation addresses both failures at once, and this is the
precise sense in which it is \emph{mandatory} rather than optional.  It is not a device
for removing a divergent force (no such force exists in the in--in equation); it is a
reorganization of the theory---an exact resummation.  The description shifts from
``mean field plus perturbative quantum fluctuations around it,'' which the infrared
divergence renders useless, to ``an ensemble of classical stochastic histories,'' each
of which is finite and physical, with the divergence relocated into the spread of the
ensemble.  Correlators computed from the Langevin dynamics are finite and
self-consistently regulated by the dissipation and the nonlinearity---the
Starobinsky--Yokoyama equilibrium distribution~\cite{StarobinskyYokoyama1994} being the
canonical example, known to reproduce the resummation of the leading infrared
logarithms to all orders.  The identity \eqref{eq:HS} guarantees consistency (no double
counting); the physics it unlocks is nonperturbative resummation and, equally
important, access to single-history questions: a physical measurement is one
realization, not an ensemble average, and only the stochastic representation describes
what one realization does.

\paragraph{Why no coarse graining is needed: self-selection of the order parameter.}
The reason the stochastic representation requires no coarse graining is worth stating
once as a principle, since it recurs in every arena below.  In a system with an
instability or an infrared divergence, a macroscopic order parameter $q$ emerges
\emph{dynamically} of its own accord: the unstable or infrared-enhanced long-wavelength
mode grows, becomes classical, and comes to dominate and characterize the whole system.
The remaining microscopic degrees of freedom do not organize into structure; they act as
an \emph{effective environment} for $q$ whether or not one chooses to call them so.  The
decomposition ``system $=$ order parameter $+$ effective environment'' is thus not an
external modeling choice imposed by the physicist, but a structure the dynamics selects
by itself---one is compelled to follow $q$ because it is what becomes macroscopic and
visible.  Coarse graining, in the usual sense, only re-describes after the fact a split
that the instability has already made.  This is why the same Hubbard--Stratonovich
identity operates identically for soft QED, soft gravitons, de Sitter fields, and phase
transitions: in each, an infrared or unstable sector self-selects the order parameter,
and the rest supplies the noise.  Throughout this paper, when we say that the
construction needs no coarse graining, this self-selection is what is meant.

The original infrared divergence is not erased by the separation; it is relocated
into the stochastic covariance.  For a $1/f$ kernel this relocation is explicit: the
equal-time variance
\begin{equation}
  \langle\xi^2\rangle
  = \frac{A}{\pi}\ln\frac{\omega_{\max}}{\omega_{\IR}}
  \;\xrightarrow{\;\omega_{\IR}\sim T_{\obs}^{-1}\;}\;
  \frac{A}{\pi}\ln(\omega_{\max}\,T_{\obs})
  \label{eq:IRDrelocated}
\end{equation}
grows logarithmically with observation time $T_{\obs}$.  The $-1$ power law persists
toward lower frequency as $T_{\obs}$ increases and diverges only in the ideal limit
$T_{\obs}\to\infty$.  This is the precise ``price'' of the original IRD: not an
infinite deterministic force, but an indefinitely accumulating stochastic variance
that is finite and well-defined at every finite observation time.

\section{The \texorpdfstring{$1/f$}{1/f} infrared sector}
\label{sec:oneoverf}

A universal long-memory kernel is obtained when the noise spectrum has the form
\begin{equation}
  N_{1/f}(\omega)
  = \frac{A}{|\omega|}\,
  \Theta(|\omega|-\omega_{\IR})\Theta(\omega_{\max}-|\omega|) .
  \label{eq:oneoverfSpectrum}
\end{equation}
The integrated variance is
\begin{equation}
  \langle \xi^2\rangle
  = \int \frac{\dd \omega}{2\pi}N_{1/f}(\omega)
  = \frac{A}{\pi}\ln \frac{\omega_{\max}}{\omega_{\IR}} ,
  \label{eq:logDivergence}
\end{equation}
which is the original IRD now residing in the stochastic covariance rather than in a
deterministic coefficient.

For an observed variable $q$ with linear response function $\chi(\omega)$
(e.g.\ a Drude conductivity), the observable power spectrum is
\begin{equation}
  S_q(\omega) = |\chi(\omega)|^2\,N_{1/f}(\omega) .
  \label{eq:Sobs}
\end{equation}
If $|\chi(\omega)|^2$ is approximately constant at low frequency, $S_q(\omega)\propto
1/|\omega|$, giving $1/f$ noise directly.

Two time scales must be distinguished.  The upper frequency $\omega_{\max}$ fixes the
shortest time in the infrared scaling regime,
\begin{equation}
  t_{\UV}=\omega_{\max}^{-1} .
\end{equation}
The total observation time fixes the lower cutoff,
\begin{equation}
  t_{\IR}=\omega_{\IR}^{-1}\sim T_{\obs} .
\end{equation}
Increasing $T_{\obs}$ lowers $\omega_{\IR}$ and enlarges the logarithm
\eqref{eq:logDivergence} while the $-1$ power law itself remains.  This is the precise
sense in which the original IRD survives: not as an infinite deterministic force, but
as the indefinitely extensible long-memory part of the stochastic source.

\section{Soft QED: \texorpdfstring{$1/f$}{1/f} current noise and source-parameter dependence}
\label{sec:qed}

\subsection{Gauge-invariant influence functional}

The QED soft sector is the cleanest arena because the photon field is Gaussian before
coupling to charged matter, while the leading soft corrections are constrained by Ward
identities and exponentiate.

Consider a macroscopic conserved current built from $N$ charged carriers, scattered for
example by impurities, with charge $e$, carrier density $n$, and characteristic velocity
$\beta=v/c$:
\begin{equation}
  j^\mu(x)=e\sum_{i=1}^{N}\int \dd\tau_i\,u_i^\mu(\tau_i)
  \delta^{(4)}(x-z_i(\tau_i)) ,
  \qquad \partial_\mu j^\mu=0 .
  \label{eq:totalCurrent}
\end{equation}
Integrating out the photon field on the CTP contour gives the influence action
\begin{equation}
  \Gamma_\gamma[j_r,j_a]
  = -\int \dd^4x\dd^4x'\,j_a^\mu(x)D^{R}_{\mu\nu}(x-x')j_r^\nu(x')
  + \frac{\ii}{2}\int \dd^4x\dd^4x'\,j_a^\mu(x)D^{H}_{\mu\nu}(x-x')j_a^\nu(x') .
  \label{eq:QEDInfluence}
\end{equation}
The Hadamard function
\begin{equation}
  D^H_{\mu\nu}(x-x')
  = \langle\{A_\mu(x),A_\nu(x')\}\rangle
  = \int\frac{\dd^4k}{(2\pi)^4}\,
  2\pi\delta(k^2)\,P^T_{\mu\nu}(k)\,e^{-\ii k(x-x')}
  \label{eq:HadamardPhoton}
\end{equation}
is positive on physical transverse modes, or equivalently on conserved currents.
Here $P^T_{\mu\nu}(k)$ is the transverse polarization sum: in a physical (radiation)
gauge with a timelike reference vector $n^\mu$ it reads
\begin{equation}
  P^T_{\mu\nu}(k)
  = \sum_{\lambda=1,2}\epsilon^{(\lambda)}_\mu(k)\,\epsilon^{(\lambda)*}_\nu(k)
  = -\,\eta_{\mu\nu}
  + \frac{k_\mu n_\nu + k_\nu n_\mu}{k\cdot n}
  - \frac{k_\mu k_\nu}{(k\cdot n)^2} ,
  \label{eq:transverseProjector}
\end{equation}
which satisfies $k^\mu P^T_{\mu\nu}=0$ on shell; when contracted with a conserved
current ($k_\mu j^\mu=0$) only the spatial transverse part $\delta_{ij}-\hat
k_i\hat k_j$ survives, so the gauge-dependent terms drop out.  The gauge-invariant form
of \eqref{eq:QEDInfluence} is a CTP Wilson-line expectation value (see
Appendix~\ref{app:wilson}), which automatically combines self-energy and vertex
infrared contributions.  By the Ward--Takahashi identity~\cite{Ward1950,Takahashi1957}
\begin{equation}
  k_\mu\Gamma^\mu(p+k,p)=S^{-1}(p+k)-S^{-1}(p) ,
  \label{eq:Ward}
\end{equation}
where $\Gamma^\mu(p+k,p)$ is the full electron--photon vertex function and $S(p)$ is the
full electron propagator (so that $S^{-1}(p)$ is the inverse propagator), the vertex and
self-energy IRDs are not independent noise channels; they constitute a single
gauge-invariant object.

\paragraph{Diagrammatic origin: a cut single-photon exchange.}
Diagrammatically, \eqref{eq:QEDInfluence} is the diagram in which one photon line
connects two insertions of the current $j^\mu$; $D^R_{\mu\nu}$ is this diagram evaluated
in the ordinary (uncut) sense, and $D^H_{\mu\nu}$ in \eqref{eq:HadamardPhoton} is the
same diagram with the internal photon line cut on shell---the real-time cutting rule of
the general discussion in Section~\ref{sec:ctp}, here reducing to the familiar
Bloch--Nordsieck/Yennie--Frautschi--Suura statement that the imaginary part of the
virtual-photon exchange equals the phase-space integral over real soft-photon emission.
Cutting a bare one-loop electron self-energy diagram alone would not be gauge invariant;
consistency requires including the cut vertex diagrams as well, which is the
content of the Ward--Takahashi identity \eqref{eq:Ward}.  The Wilson-line representation
of Appendix~\ref{app:wilson} sidesteps this bookkeeping: since it contains only
insertions along a single worldline, with no distinction between ``self-energy'' and
``vertex'' attachment points, its cut is automatically the gauge-invariant sum of all
such contributions to every order, and reduces to \eqref{eq:HadamardPhoton} directly
(see Appendix~\ref{app:wilson} for the explicit consistency check between the resummed
exponent and \eqref{eq:Ward}).

\subsection{Noise kernel and \texorpdfstring{$1/f$}{1/f} spectrum}
\label{sec:qed-noise}

The Hubbard--Stratonovich identity applied to \eqref{eq:QEDInfluence} produces a
$c$-number stochastic gauge potential $a_\mu$ with
$\langle a_\mu(x)a_\nu(x')\rangle = D^H_{\mu\nu}(x-x')$.
For the current \eqref{eq:totalCurrent}, the current-projected noise kernel in the
soft limit ($|\bm{k}|\to 0$) follows from the on-shell photon phase space,
\begin{equation}
  \int \frac{\dd^3k}{(2\pi)^3\,2|\bm{k}|}\,
  \frac{(p\cdot\epsilon)^2}{(p\cdot k)^2}
  \;\propto\;
  \int_{\omega_{\IR}}^{\omega_{\max}} \frac{\dd\omega}{\omega}\,
  f(\beta) ,
  \label{eq:softPhaseSpace}
\end{equation}
with a velocity form factor such as
\begin{equation}
  f(\beta)=\frac{1}{\beta}\ln\frac{1+\beta}{1-\beta}-2
  \xrightarrow{\beta\to 0} \frac{2}{3}\beta^2,\qquad
  f(\beta)\xrightarrow{\beta\to 1} \infty .
  \label{eq:velFactor}
\end{equation}
For $N=nV$ identical sources with isotropic velocity distribution, the angular integral
and sum over sources give the noise kernel
\begin{equation}
  N^{\mu\nu}_{\rm QED}(\omega)
  = n\,e^2\,
  \frac{f(\beta)}{|\omega|}\,
  P^T_{\mu\nu}\,
  \Theta(|\omega|-\omega_{\IR})\Theta(\omega_{\max}-|\omega|)
  \label{eq:QEDnoiseKernel}
\end{equation}
(see Appendix~\ref{app:wilson} for the detailed derivation).

This is a central result.  The noise spectrum is $N(\omega)\propto 1/|\omega|$, i.e.\
$1/f$, arising purely from the soft-photon phase space.  The amplitude is determined
entirely by Ward-identity-constrained source parameters: $n$, $e$, $\beta$.  In the
non-relativistic limit $\beta\ll 1$, the amplitude scales as $n\,e^2\beta^2$.

It is worth being explicit about what $N^{\mu\nu}_{\rm QED}$ represents physically,
since three related but distinct two-point objects appear in this section and the next.
The Hubbard--Stratonovich field $a_\mu$ has the correlator $\langle a_\mu a_\nu\rangle
= D^H_{\mu\nu}$ \eqref{eq:HadamardPhoton}: this is the statistical correlation of the
random gauge field at the level of the photon field itself, computed here in vacuum,
without any scattering medium.  Projecting this onto the charged current
\eqref{eq:totalCurrent}---contracting with the carrier four-velocities and integrating
over the on-shell soft-photon phase space---yields $N^{\mu\nu}_{\rm QED}(\omega)$, which
is the spectrum of the stochastic \emph{force} acting on the current, i.e.\ the
covariance of the source $\xi$ on the right-hand side of the Langevin equation
\eqref{eq:LangevinKernel}.  It is not yet the spectrum of the current fluctuation
itself.  The latter, $S_J(\omega)$, follows only after the force is propagated through
the linear response (susceptibility) of the medium; this is the step taken in
Section~\ref{sec:qed-drude}, where the impurity scattering that defines that response
first enters.  Schematically,
\begin{equation}
  \underbrace{D^H_{\mu\nu}}_{\text{photon-field noise}}
  \;\xrightarrow{\ \text{project on current}\ }\;
  \underbrace{N^{\mu\nu}_{\rm QED}}_{\text{force on the current}}
  \;\xrightarrow{\ \times\,|\chi(\omega)|^2\ }\;
  \underbrace{S_J(\omega)}_{\text{current fluctuation}} .
  \label{eq:threekernels}
\end{equation}

\subsection{Stochastic equation of motion and impurity-density dependence}
\label{sec:qed-drude}

We now propagate the stochastic force of the previous section through a conducting
medium, where impurity scattering supplies both the dissipation and the upper edge of
the scaling band.  For an impurity-dominated conductor, the drift velocity of a charge
carrier obeys the Drude--Langevin equation
\begin{equation}
  m\dot v(t)+m\gamma_{\rm imp}v(t)=eE+\xi_{\rm IR}(t)+\xi_{\rm th}(t),
  \label{eq:DrudeLangevin}
\end{equation}
where $\xi_{\rm th}$ is thermal (Johnson--Nyquist) noise and
\begin{equation}
  \gamma_{\rm imp}=\tau_{\rm imp}^{-1}\simeq n_{\rm imp}v_F\sigma_{\rm tr}
  \label{eq:gammaImp}
\end{equation}
is the impurity scattering rate.  The infrared force $\xi_{\rm IR}$ is the soft-photon
stochastic source of Section~\ref{sec:qed-noise}, the scalar projection of
$N^{\mu\nu}_{\rm QED}$ onto the current direction; its stationary correlation is
\begin{equation}
  \langle \xi_{\rm IR}(t)\,\xi_{\rm IR}(t')\rangle
  = \int\frac{\dd\omega}{2\pi}\,e^{-\ii\omega(t-t')}\,N_{\rm IR}(\omega),
  \qquad
  N_{\rm IR}(\omega) = \frac{C_{\rm IR}(n_{\rm imp})}{|\omega|},
  \label{eq:NIRdef}
\end{equation}
the $1/|\omega|$ form following from the soft-photon phase space
\eqref{eq:QEDnoiseKernel}; a scaling estimate of the coefficient $C_{\rm IR}$ is given
in Appendix~\ref{app:CIR}.

We want the power spectral density of the electric current.  For carrier density $n$
the current density is $J=-nev$, and we define its spectral density $S_J(\omega)$ in the
stationary state by the Wiener--Khinchin relation
\begin{equation}
  \langle \delta J(t)\,\delta J(t')\rangle
  = \int\frac{\dd\omega}{2\pi}\,e^{-\ii\omega(t-t')}\,S_J(\omega),
  \qquad
  S_J(\omega) = (ne)^2\,S_v(\omega),
  \label{eq:SJdef}
\end{equation}
where $S_v(\omega)$ is the velocity spectral density and $\delta J=-ne\,\delta v$ is the
fluctuating part of the current.  Solving \eqref{eq:DrudeLangevin} in frequency space
for the fluctuating velocity (dropping the deterministic drive $E$, which sets the mean
current rather than its fluctuations),
\begin{equation}
  \delta v(\omega)
  = \frac{\xi_{\rm IR}(\omega)+\xi_{\rm th}(\omega)}{m(\gamma_{\rm imp}-\ii\omega)},
  \label{eq:vsolution}
\end{equation}
the response is the Drude (Lorentzian) susceptibility
$\chi(\omega)=[m(\gamma_{\rm imp}-\ii\omega)]^{-1}$, with
$|\chi(\omega)|^2=[m^2(\gamma_{\rm imp}^2+\omega^2)]^{-1}$.  Taking the modulus squared
of \eqref{eq:vsolution} and using the stationarity and independence of the two noises,
\begin{equation}
  S_v(\omega)
  = |\chi(\omega)|^2\big[N_{\rm IR}(\omega)+N_{\rm th}(\omega)\big]
  = \frac{N_{\rm IR}(\omega)+N_{\rm th}(\omega)}
         {m^2(\gamma_{\rm imp}^2+\omega^2)}.
  \label{eq:Sv}
\end{equation}
Retaining the infrared part and evaluating it in the low-frequency band
$|\omega|\ll\gamma_{\rm imp}$, where $|\chi(\omega)|^2\simeq(m\gamma_{\rm imp})^{-2}$ is
flat, the current spectral density \eqref{eq:SJdef} becomes
\begin{equation}
  S_J^{\rm IR}(\omega)
  = (ne)^2\,|\chi(\omega)|^2\,N_{\rm IR}(\omega)
  \simeq \left(\frac{ne}{m\gamma_{\rm imp}}\right)^2
  \frac{C_{\rm IR}(n_{\rm imp})}{|\omega|}
  \equiv \frac{A_J(n_{\rm imp})}{|\omega|}
  \label{eq:SJmain}
\end{equation}
with the $1/f$ noise coefficient
\begin{equation}
  A_J(n_{\rm imp})
  = \left(\frac{ne}{m\gamma_{\rm imp}}\right)^2 C_{\rm IR}(n_{\rm imp}) .
  \label{eq:AJmain}
\end{equation}
Thus the flat Drude response converts the $1/|\omega|$ force kernel directly into a
$1/f$ current spectrum; the Lorentzian rolloff at $|\omega|\gtrsim\gamma_{\rm imp}$ fixes
the upper edge of the scaling band, $\omega_{\max}\sim\gamma_{\rm imp}$.  If $C_{\rm
IR}\propto n_{\rm imp}$ (one soft-dressing event per scattering) and $\gamma_{\rm
imp}\propto n_{\rm imp}$, then $A_J\propto n_{\rm imp}^{-1}$.  This is a falsifiable
scaling prediction.

Comparing with the empirical Hooge formula~\cite{Hooge1969},
\begin{equation}
  \frac{S_I(f)}{I^2}=\frac{\alpha_H}{N_c f} ,
  \label{eq:Hooge}
\end{equation}
the coefficient $A_J\propto C_{\rm IR}/\gamma_{\rm imp}^2$ of the \emph{absolute}
spectrum $S_J=A_J/|\omega|$ is calculable from the microscopic soft-photon influence
functional, and provides a first-principles estimate of the otherwise phenomenological
Hooge parameter.  For example, identifying the two spectra in the flat Drude band gives,
up to the normalization by $I^2$ and the carrier number $N_c$, $\alpha_H\sim C_{\rm
IR}/\gamma_{\rm imp}^{\,p}$ with an exponent $p$ fixed by that normalization; the
resulting dependence on impurity density and mobility is a falsifiable prediction, to be
tested against samples with controlled $n_{\rm imp}$.  We keep the absolute/relative
distinction explicit because the power of $\gamma_{\rm imp}$ in $\alpha_H$ is sensitive
to the $I^2$ and $N_c$ normalization, whereas the absolute-spectrum result
$A_J\propto C_{\rm IR}/\gamma_{\rm imp}^2$ is not.  The thermal contribution $\xi_{\rm
th}$ gives the standard Johnson--Nyquist white noise $S_J^{\rm th}\propto T\gamma_{\rm
imp}$~\cite{Johnson1928,Nyquist1928,CallenWelton1951}, which is distinct from and
additive to the $1/f$ term.

The memory interpretation is also direct: soft photons carry which-path information
about charged-particle histories.  The integrated $1/f$ variance
\begin{equation}
  \langle\delta J^2\rangle_{\rm IR}
  \simeq \frac{A_J}{\pi}\ln\frac{\omega_{\max}}{\omega_{\IR}}
\end{equation}
is the electrical analogue of the inflationary variance $\langle\delta\phi^2\rangle
\simeq (H/2\pi)^2\ln(k_{\max}/k_{\min})$.

\paragraph{Phenomenological verification with electron wave packets.}
The abstract prediction of a $1/f$ current noise from the soft-photon back reaction
has been verified in a concrete, phenomenological wave-packet model in a companion
study~\cite{MorikawaNakamichi2023pink}.  There the electron is represented by a
classical wave packet propagating in a (semi-)conductor and obeying a classical
Langevin equation that includes the soft-photon back reaction; squaring the wave packet
to form the electric current and demodulating the result yields a robust $1/f$ spectrum.
That treatment complements the field-theoretic derivation given here: the present paper
fixes the noise amplitude and its parameter dependence from the CTP influence
functional, while the wave-packet model exhibits the robustness and propagation of the
resulting $1/f$ signal and traces it through to thresholded, demodulated output.  The
two descriptions share the same physical origin---the soft-photon infrared sector
acting back on the charged current---and are mutually consistent: the demodulation
(beat) picture~\cite{MorikawaNakamichi2023amplitude} and the influence-functional
picture are complementary readings of the same infrared structure.

\section{Cross sections versus order parameters: a no-go objection and its scope}
\label{sec:fukuyama}

The construction of the previous section assigns a stochastic interpretation to the
soft sector of QED, a step that has recently been challenged on careful and physically
motivated grounds by Fukuyama~\cite{Fukuyama2025}.  Because the objection is precise
and instructive, we address it in detail.  We will argue that the no-go result is
correct for the class of observables it considers---inclusive scattering cross
sections of the free or source-free theory---and that the order-parameter dynamics of
this paper lies outside that class.  The two analyses are therefore complementary
rather than contradictory, and drawing the boundary between them clarifies precisely
when an infrared sector does and does not classicalize.

\subsection{The two claims of the no-go argument}

Fukuyama's analysis rests on two claims, both of which we accept within their stated
scope.

\paragraph{Claim 1 (inclusive cancellation).}
In the Fock-space formulation, the soft-photon phases accompanying a charged particle
are fixed by gauge invariance through the Ward--Takahashi identity.  (The eikonal factor
is defined precisely in Appendix~\ref{app:wilson}: the universal, spin-independent
vertex $eu^\mu/(u\cdot k)$ that arises when a soft photon of momentum $k$ attaches to a
charge moving with velocity $u^\mu$ along an undeflected classical trajectory.)  For any
\emph{inclusive} observable---a cross section summed over unresolved final
states---the virtual and real soft contributions carry the same eikonal factor $L$ and
cancel by the Bloch--Nordsieck/KLN mechanism.  The infrared sector is then a coherent
unitary dressing, $|\Psi\rangle\to e^{\ii\Phi_{\rm IR}}|\Psi\rangle$, and its
fluctuations are Callen--Welton quantum fluctuations~\cite{CallenWelton1951}, not
classical noise.

\paragraph{Claim 2 (conformal protection).}
The four-dimensional Maxwell action is conformally invariant.  Consequently the
\emph{free} electromagnetic field has no infrared growth in a de Sitter background:
long-wavelength photon modes do not freeze, and no secular accumulation of infrared
power occurs.  The imaginary part of the Schwinger--Keldysh effective action therefore
does not grow, and although a Hubbard--Stratonovich auxiliary field can still be
introduced formally, it does not acquire the dynamical status of a Langevin force.

\subsection{Why neither claim constrains order-parameter dynamics}

\paragraph{The observable is not a cross section.}
Claim 1 is a statement about inclusive cross sections: quantities of the form
$\sum_f|\mathcal{M}_{i\to f}|^2$, in which the sum over final soft-photon
configurations is precisely what produces the BN/KLN cancellation.  The observable of
the present paper is of a different type.  It is the real-time two-point function of a
collective order parameter,
\begin{equation}
  S_J(\omega)=\int \dd t\,e^{\ii\omega t}\,
  \langle\,\delta J(t)\,\delta J(0)\,\rangle ,
  \label{eq:JJcorr}
\end{equation}
the symmetrized current--current correlator.  This is not an inclusive sum over final
states; it is a fixed-operator expectation value in a definite state, and there is no
KLN theorem that forces its infrared part to cancel.  Indeed, the very mechanism that
cancels the inclusive cross section---summing $|\mathcal{M}|^2$ over emitted
photons---does not act on \eqref{eq:JJcorr}, because $\delta J(t)$ is a single
Heisenberg operator, not a transition probability.  The infrared content of
\eqref{eq:JJcorr} is the current-projected symmetric kernel
\eqref{eq:QEDnoiseKernel}---the photon Hadamard function \eqref{eq:HadamardPhoton}
projected onto the conserved current, following the hierarchy \eqref{eq:threekernels}---which
is positive and does not cancel against the antisymmetric (commutator) part that
supplies the retarded response.

\paragraph{From a single dressing to an ensemble: the role of memory.}
Claim 1 treats the soft cloud as a single unitary dressing $U_{\rm IR}$, valid for a
single scattering event.  Fukuyama himself contrasts the unitary evolution
$\rho\to U_{\rm IR}\rho U_{\rm IR}^\dagger$ (his Eq.~34) with the stochastic ensemble
$\rho\to\int d\xi\,P(\xi)\,U(\xi)\rho\,U(\xi)^\dagger$ (his Eq.~33).  For a single event
the former is correct.  But an order parameter is not a single event: the current
$J(t)$ integrates over a continuous history of scattering events, each updating the
soft dressing by a different amount $U_{\rm IR}[J(t)]$.  Averaging over the unresolved
soft history of this \emph{sequence} of dressings is just the passage from Eq.~(34)
to Eq.~(33).  The object that records this accumulation is the memory: the retarded
kernel $\Sigma_R(t,t')$ of the CTP effective action, represented phenomenologically by
the damping term in \eqref{eq:DrudeLangevin}, stores the influence of past dressings on
present dynamics.  A single inclusive cross section, having no time extension, carries
no such memory; the order parameter does.  This is why the stochastic representation
appears for the latter and not the former, with no contradiction between them.

\paragraph{Conformal protection applies to the dry, source-free sector.}
Claim 2 concerns the \emph{free} Maxwell field, whose conformal invariance forbids
infrared growth.  In the language of Section~\ref{sec:ds}, this is precisely a
\emph{dry} noise: a kinematical fluctuation with no interaction-generated retarded
partner.  We fully agree that a dry electromagnetic noise does not classicalize---this
is the gauge-field counterpart of our statement that free de Sitter scalar noise is dry
and that interaction is required for a genuine stochastic dynamics.  The construction of
Section~\ref{sec:qed}, however, is not the free Maxwell field.  It is the soft sector
\emph{coupled to a conserved charged current} of massive particles with density $n$ and
scattering rate $\gamma_{\rm imp}$.  The electron mass $m$ and the scattering scale
$\gamma_{\rm imp}$ break conformal invariance explicitly: they introduce the very scales
$\omega_{\max}\sim\gamma_{\rm imp}$ and $t_{\UV}=\omega_{\max}^{-1}$ that define the
infrared band of \eqref{eq:QEDnoiseKernel}.  The $1/f$ spectrum does not rely on
de Sitter secular growth; it arises in flat spacetime from the soft-photon phase
space of an interacting current, with the lower edge $\omega_{\IR}\sim T_{\obs}^{-1}$
set by observation time.  Conformal protection of the free field and $1/f$ noise of the
interacting current are simply statements about two different theories.

\subsection{A unified reading}

Far from undermining the present framework, Fukuyama's no-go result fits neatly into
it.  His Claim 2 is the QED instance of the dry/non-dry distinction: the conformally
protected free field is the dry sector that does not classicalize as we
require.  His Claim 1 correctly identifies that inclusive cross sections are
infrared-finite and coherent; we add only that the complementary real-time
order-parameter correlator is infrared-sensitive and carries the memory and noise that
inclusive sums discard.  The distinction between Callen--Welton quantum fluctuations
and classical noise, on which Claim 1 turns, is itself quantified by the occupation
criterion \eqref{eq:occupationRatio}: a fluctuation is irreducibly quantum where the
occupation ratio is of order one, and asymptotically classical where it diverges.  The
soft sector of an evolving current sits in the second regime---its occupation
$\langle N_\gamma\rangle\sim\int\dd\omega/\omega$ diverges in the infrared---so the
boundary between Callen--Welton fluctuation and classical noise is not crossed by fiat
but by the infrared divergence itself.  The boundary Fukuyama draws---between coherent
inclusive amplitudes and irreversible classicalized dynamics---is the same boundary we
draw between the retarded (coherent, causal) and symmetric (noise) parts of the CTP
effective action.  We are indebted to his analysis for making the location of that
boundary explicit.  The disagreement, to the extent there is one, is only over which
side of it contains the electric current.  Because the current is an order parameter
with real-time memory rather than an inclusive cross section, it belongs on the
stochastic side.

There is in fact no tension with Fukuyama's emphasis on coherent dressing.  The
Kulish--Faddeev dressing is precisely the coherent state that a classical current
produces when it drives the photon field (Appendix~\ref{app:wilson}).  A single dressing
is coherent, as Fukuyama stresses; but an order parameter evolving in real time drives a
\emph{succession} of such coherent states, and the ensemble average over their
unresolved soft history is the stochastic kernel.  Coherent dressing and stochastic noise
are thus the static and dynamic faces of the same current-driven coherent-state physics.

That soft radiation carries genuine, physical which-path information---and is not a mere
coherent phase to be gauged away---is independently supported by recent work outside the
$1/f$ context.  Danielson, Satishchandran, and Wald~\cite{DanielsonSatishchandranWald2023}
show that a black-hole (or Killing) horizon decoheres a quantum superposition through the
emission of soft photons and gravitons that imprint a which-path memory, at a rate set by
the soft-emission phase space.  This is the same soft-sector physics that, in the present
real-time setting, supplies the noise kernel.  In the $S$-matrix framework itself, Hirai
and Sugishita~\cite{HiraiSugishita2023} have shown that interference effects between
different hard processes cannot be computed with Fock states at all---the infrared
divergence sets them to zero---and require appropriately dressed initial and final
states; the soft sector must be treated as physical before even the interference
content of scattering theory is well defined.  These results corroborate the reading
advanced here: the soft sector is physically active in real-time, memory-carrying
processes, even though it cancels in inclusive cross sections.

\section{Soft gravitons, metric noise, and gravitational memory}
\label{sec:gravity}

\subsection{Gravitational influence functional}

Gravity has an infrared structure parallel to QED but governed by energy-momentum
rather than charge.  We construct it in analogy with Section~\ref{sec:qed}.
The parallel is: photon field $A_\mu\to$ graviton field $h_{\mu\nu}$ (the metric
perturbation around a background $g_{\mu\nu}$), conserved current $j^\mu\to$ stress
tensor $T^{\mu\nu}$, and the Ward--Takahashi identity $\to$
diffeomorphism (stress-energy conservation).

Just as we projected the photon Hadamard kernel $D^H_{\mu\nu}$ onto the current, we
project the graviton Hadamard kernel onto the stress tensor.  The graviton field has the
Hadamard (symmetric) two-point function
\begin{equation}
  \mathcal{D}^H_{\mu\nu\alpha\beta}(x-x')
  = \langle\{h_{\mu\nu}(x),h_{\alpha\beta}(x')\}\rangle
  = \int\frac{\dd^4k}{(2\pi)^4}\,2\pi\delta(k^2)\,
  \Lambda_{\mu\nu\alpha\beta}(k)\,e^{-\ii k(x-x')} ,
  \label{eq:gravHadamard}
\end{equation}
with $\Lambda_{\mu\nu\alpha\beta}$ the transverse-traceless projector defined in
\eqref{eq:TTprojector}.  This is the spin-2 counterpart of the photon Hadamard kernel
\eqref{eq:HadamardPhoton}; like it, it is a regular distribution and carries no infrared
divergence by itself.  The metric perturbation $h_{\mu\nu}$ couples to the stress tensor
through $\tfrac{1}{2}\int h^{\mu\nu}T_{\mu\nu}$, so separating the soft graviton sector
and projecting \eqref{eq:gravHadamard} onto the stress-energy source gives a
closed-time-path influence functional for $h_{\mu\nu}$,
\begin{align}
  \Gamma_g[h_r,h_a]
  &= \int \dd^4x\,h_a^{\mu\nu}
    \left[G^{(1)}_{\mu\nu}[h_r]-8\pi G\langle\hat{T}_{\mu\nu}\rangle\right]
  \nonumber\\
  &\quad +\int \dd^4x\dd^4x'\,
    h_a^{\mu\nu}(x)\,\mathcal{H}_{\mu\nu\alpha\beta}(x,x')\,h_r^{\alpha\beta}(x')
  \nonumber\\
  &\quad +\frac{\ii}{2}\int \dd^4x\dd^4x'\,
    h_a^{\mu\nu}(x)\,\mathcal{N}_{\mu\nu\alpha\beta}(x,x')\,h_a^{\alpha\beta}(x')
  +\cdots ,
  \label{eq:gravInfluence}
\end{align}
whose imaginary part is the stress-tensor noise kernel
\begin{equation}
  \mathcal{N}_{\mu\nu\alpha\beta}(x,x')
  = \frac{1}{2}\langle\{\hat{t}_{\mu\nu}(x),\hat{t}_{\alpha\beta}(x')\}\rangle ,
  \qquad
  \hat{t}_{\mu\nu}=\hat{T}_{\mu\nu}-\langle\hat{T}_{\mu\nu}\rangle .
  \label{eq:stressNoise}
\end{equation}
The factor $\tfrac12$ in \eqref{eq:stressNoise} follows the standard stochastic-gravity
normalization~\cite{HuVerdaguer2008}, whereas the photon and graviton Hadamard functions
\eqref{eq:HadamardPhoton} and \eqref{eq:gravHadamard} were defined without it; the
difference is a bookkeeping convention that cancels in every observable spectrum, which
depends only on the product of kernel and response.
This is the source-projected form of \eqref{eq:gravHadamard}, just as
$N^{\mu\nu}_{\rm QED}$ was the current-projected form of $D^H_{\mu\nu}$; the infrared
divergence appears only upon this projection and integration over the soft-graviton
phase space, not in \eqref{eq:gravHadamard} itself.  It is positive by construction, and
diffeomorphism invariance requires
\begin{equation}
  \nabla^\mu \mathcal{N}_{\mu\nu\alpha\beta}(x,x')=0 ,
  \label{eq:transversality}
\end{equation}
the gravitational analogue of the Ward--Takahashi identity \eqref{eq:Ward}.  As in the
QED case, \eqref{eq:gravInfluence} is diagrammatically a single graviton line exchanged
between two stress-tensor insertions: $\mathcal H_{\mu\nu\alpha\beta}$ is this diagram
uncut, and $\mathcal N_{\mu\nu\alpha\beta}$ is the same diagram with the internal
graviton line cut on shell, $2\pi\delta(k^2)\Lambda_{\mu\nu\alpha\beta}$, the direct
gravitational analogue of the real-time cutting rule discussed in
Section~\ref{sec:ctp}.  Stress-energy conservation \eqref{eq:transversality} plays the
role that the Ward--Takahashi identity plays in QED, guaranteeing that this cut is
transverse and hence physical; it is the diagrammatic underpinning of the universality
of Weinberg's soft-graviton theorem, just as \eqref{eq:Ward} underlies the universality
of the soft-photon factor.

The same influence functional and its Einstein--Langevin consequence below were obtained
by Hu and Verdaguer in the stochastic-gravity
programme~\cite{HuVerdaguer2008,CalzettaHu2008} by treating the quantum matter fields as
an environment.  As explained in Section~\ref{sec:ctp}, that step is a re-description
rather than a necessity: the noise kernel \eqref{eq:stressNoise} is the soft-graviton
Hadamard kernel projected onto the stress-energy source, and its Hubbard--Stratonovich
separation is an identity of the generating functional.  The unstable gravitational
infrared sector self-selects the metric perturbation as the order parameter.

\subsection{Einstein--Langevin equation}

Applying the Hubbard--Stratonovich identity to \eqref{eq:gravInfluence} yields a
$c$-number stochastic metric source $\xi_{\mu\nu}$ with
$\langle\xi_{\mu\nu}\xi_{\alpha\beta}\rangle = \mathcal{N}_{\mu\nu\alpha\beta}$, and
the Einstein--Langevin equation~\cite{HuVerdaguer2008},
\begin{equation}
  G^{(1)}_{\mu\nu}[h]
  -8\pi G\langle\hat{T}_{\mu\nu}\rangle
  +\int \dd^4x'\,\mathcal{H}_{\mu\nu\alpha\beta}(x,x')\,h^{\alpha\beta}(x')
  = 8\pi G\,\xi_{\mu\nu}(x) .
  \label{eq:EinsteinLangevin}
\end{equation}
This is not a phenomenological addition to semiclassical gravity: it is the CTP-derived
representation obtained by separating the soft graviton sector, in complete parallel with
the QED Langevin equation \eqref{eq:DrudeLangevin}.

\subsection{Soft gravitons, \texorpdfstring{$1/f$}{1/f} metric noise, and memory}

From high-energy scattering, the Weinberg soft
theorem~\cite{Weinberg1965,Strominger2018} gives the leading soft-graviton emission
amplitude as
\begin{equation}
  S_g(k)=\kappa\sum_i\eta_i\,
  \frac{p_i^\mu p_i^\nu\,\epsilon_{\mu\nu}(k)}{p_i\cdot k} ,
  \qquad \kappa^2=32\pi G ,
  \label{eq:softGravFactor}
\end{equation}
which exponentiates in inclusive quantities.  On the CTP contour the exponentiated
object is the stress-tensor influence functional \eqref{eq:gravInfluence}.  In the
soft limit the noise kernel behaves as
\begin{equation}
  \mathcal{N}^{\rm soft}_{\mu\nu\alpha\beta}(\omega)
  \propto \frac{G\,E^2\,f_g(\beta)}{|\omega|}\,
  \Lambda_{\mu\nu\alpha\beta}(\hat{n}) ,
  \label{eq:gravNoise}
\end{equation}
the gravitational analogue of \eqref{eq:QEDnoiseKernel}: the charge $e^2$ is replaced by
$G E^2$ (the gravitational ``charge'' is energy), the transverse projector by the
transverse-traceless one, and the dipole velocity factor $f(\beta)\sim\beta^2$ by the
quadrupole one $f_g(\beta)\sim\beta^4$.  The derivation parallels
Appendix~\ref{app:wilson} and is given in Appendix~\ref{app:graviton}.

\paragraph{Gravitational memory as the retarded kernel.}
The gravitational memory effect~\cite{Christodoulou1991,Favata2010}---the permanent
displacement of test masses after a burst of radiation---is encoded in the retarded
kernel $\mathcal{H}_{\mu\nu\alpha\beta}$ evaluated at $\omega\to 0$.  In the CTP
language, memory and stochastic metric fluctuations are the causal and symmetric parts
of the same influence functional.  An IRD in the gravitational sector therefore
simultaneously implies $1/f$ metric noise \emph{and} a long-memory retarded kernel.
The two faces of the same infrared singularity are inseparable.  This is consistent with
the infrared triangle~\cite{Strominger2018}: Strominger and
Zhiboedov~\cite{StromingerZhiboedov2016} established that gravitational memory is a
transition between degenerate BMS vacua and that its Fourier transform reproduces the
soft-graviton theorem.  In the present language, the $\omega\to0$ retarded kernel
\emph{is} the memory (the vacuum shift), while the symmetric kernel at finite $\omega$
is the noise (the uncertainty in the shift); the soft theorem relates the two.

Dressed gravitational scattering states have overlaps~\cite{Carney2017,Carney2018}
\begin{equation}
  D_{ij}^{\rm grav}\sim
  \exp\!\left[-\frac{1}{2}B_{ij}^{\rm grav}\ln\frac{\omega_{\max}}{\omega_{\IR}}
  +\ii\Phi_{ij}^{\rm grav}\right] ,
\end{equation}
where the imaginary phase encodes gravitational memory and the real exponent gives
reduced-state decoherence.  The degree of this decoherence is generically partial and
depends on the scattering kinematics and on the resolution
scale~\cite{GomezLetschkaZell2018a,GomezLetschkaZell2018b}, exactly as the
finite-observation-time logarithm here keeps $D_{ij}$ finite until the strict infrared
limit is taken.  The Anderson-orthogonality-like catastrophe of the
gravitational soft sector is therefore the gravitational counterpart of the QED
soft-sector decoherence discussed in Appendix~\ref{app:Handel}.

\section{Non-equilibrium phase transitions and order-parameter dynamics}
\label{sec:phase}

\subsection{Setting}

A phase transition is the clearest realization of the self-selection principle of
Section~\ref{sec:ctp}: the order parameter does not exist as a classical variable before
the transition, yet emerges as the dominant dynamical entity after it, with the remaining
modes acting as its effective environment.  Following the original in-in treatment
of~\cite{Morikawa1995,Morikawa1986}, the dissipation and noise are induced by the
radiative corrections of the \emph{self-interacting unstable field itself}, not by
tracing out an external bath.  (The unstable field was one of the three systems for which
a Langevin equation was first derived from the CTP effective action
in~\cite{Morikawa1986}, alongside the thermal field and cosmological particle creation;
the phase-transition case was developed in~\cite{Morikawa1995}.)

Consider an order-parameter field $\Phi$ with an unstable (double-well) potential
$U(\Phi)$,
\begin{equation}
  S[\Phi]
  = \int \dd^4x\!\left[
    \tfrac{1}{2}(\partial\Phi)^2 - U(\Phi)
  \right] ,
  \qquad
  U(\Phi)=-\tfrac{1}{2}\mu^2\Phi^2+\tfrac{\lambda}{4!}\Phi^4 ,
  \label{eq:phaseAction}
\end{equation}
symmetric at high temperature and unstable (spinodal) below the critical temperature.
In parallel with the de Sitter scalar of Section~\ref{sec:ds}, the self-interaction
$\lambda\Phi^4$ generates, through its own in-in radiative corrections, both a retarded
memory kernel and a positive noise kernel; the long-wavelength (infrared) modes of
$\Phi$ that go unstable are the source of the divergence, and the retarded and symmetric
kernels are two projections of the same self-energy.  (Additional microscopic fields
$\chi$ coupled to $\Phi$ can be included; they contribute radiative corrections of the
same structure but are not required for the mechanism.)

\subsection{CTP effective action and noise kernel}

The infrared part of the effective action has the standard CTP form
\begin{align}
  \Gamma_{\rm PT}[\Phi_r,\Phi_a]
  &= \int \dd^4x\,\Phi_a\!\left[\Box\Phi_r+U'_{\eff}(\Phi_r)\right]
  \nonumber\\
  &\quad +\int \dd^4x\dd^4x'\,\Phi_a(x)\,\Sigma^R_{\rm PT}(x,x')\,\Phi_r(x')
  \nonumber\\
  &\quad +\frac{\ii}{2}\int \dd^4x\dd^4x'\,\Phi_a(x)\,N_{\rm PT}(x,x')\,\Phi_a(x') .
  \label{eq:phaseGamma}
\end{align}
The noise kernel is generated by the self-interaction of the order-parameter field, with
the same two-loop (sunset) structure as the de Sitter kernel
\eqref{eq:dSNoiseKernel}~\cite{Morikawa1995,GleiserRamos1994},
\begin{equation}
  N_{\rm PT}(x,x')
  \sim \frac{\lambda^2}{6}\,[G^H_{\rm IR}(x,x')]^3 + \cdots ,
  \label{eq:phaseNoise}
\end{equation}
$G^H_{\rm IR}$ being the Hadamard function of the infrared modes.  The retarded kernel
$\Sigma^R_{\rm PT}$ is generated by the same $\lambda\Phi^4$ interaction; both are
projections of a single self-energy.  In the cutting-rule language of
Section~\ref{sec:ctp}, \eqref{eq:phaseNoise} is the simultaneous cut of all three
internal propagators of the two-loop sunset self-energy diagram generated by the quartic
vertex, the multi-particle generalization of the single on-shell cut that produced
$N^{\mu\nu}_{\rm QED}$ and $\mathcal N_{\mu\nu\alpha\beta}$ in
Sections~\ref{sec:qed} and~\ref{sec:gravity}: one cut line per power of the coupling
that connects the two vertices, three here in place of one.  In the broken phase, where
$\Phi$ acquires a mean value $v$, expanding $U(\Phi)$ around it, $\Phi=v+\phi$,
generates an effective cubic vertex $\tfrac{\lambda v}{6}\phi^3$ for the fluctuation
$\phi$, and two such vertices form a \emph{one-loop} bubble with two internal lines
(one external leg of each cubic vertex being taken by $\Phi_a$).  Its double cut gives
an additional noise contribution
\begin{equation}
  N_{\rm PT}^{\rm mult}(x,x';\Phi_r)
  \sim \lambda^2\,\Phi_r(x)\Phi_r(x')\,[G^H_{\rm IR}(x,x')]^2 ,
  \label{eq:phaseNoiseMult}
\end{equation}
which is multiplicative---its amplitude follows the order parameter---and, being of
lower loop order, typically dominates over the $\Phi_r$-independent sunset term
\eqref{eq:phaseNoise} once $v\neq0$.  The two contributions coexist: the additive
sunset noise persists through the transition, while the multiplicative bubble noise
switches on with symmetry breaking, $N_{\rm PT}\to N_{\rm PT}(x,x';\Phi_r)$.

\subsection{Order-parameter Langevin equation}

\begin{equation}
  \Box\Phi(x)+U'_{\eff}(\Phi(x))
  +\int \dd^4x'\,\Sigma^R_{\rm PT}(x,x')\,\Phi(x')
  = -\xi_{\rm PT}(x) ,
  \label{eq:phaseLangevin}
\end{equation}
with $\langle\xi_{\rm PT}(x)\,\xi_{\rm PT}(x')\rangle = N_{\rm PT}(x,x')$.

Several features are notable.  The noise is colored (non-Markovian), and multiplicative
in the broken phase; white additive noise is at best an approximation near the critical
point.  The noise and dissipation kernels are generated by the same self-interaction,
connected by the non-equilibrium fluctuation--dissipation relation.  Near the critical
point, Goldstone and critical modes contribute infrared-enhanced noise $N_{\rm
PT}(\omega)\propto 1/|\omega|^\alpha$, a $1/f$-like spectrum whose exponent depends on
the universality class~\cite{Goldenfeld1992,ChaikinLubensky1995}.

Macroscopic classicality is not imposed: it emerges when the unstable infrared sector
is separated from the deterministic evolution and drives the system to a particular
broken-symmetry branch.  The same logic applies to cosmic inflation as a dynamical
phase transition~\cite{Morikawa2022}.

\section{Self-interacting scalar field in de Sitter space (summary)}
\label{sec:ds}

The de Sitter case has been worked out in full detail in~\cite{Morikawa2022}, where the
$\lambda\phi^4$ CTP effective action is derived at two-loop order and the
Harrison--Zel'dovich spectrum $n_s\simeq 1$ is reproduced.  The CTP treatment of de
Sitter density fluctuations goes back to~\cite{Morikawa1987}; we summarize the essential
structure here for completeness.

A light scalar in de Sitter space with $V(\phi)=\tfrac{\lambda}{4!}\phi^4+\tfrac{1}{2}
m^2\phi^2$ develops an infrared-enhanced long-wavelength sector.  The free-theory
horizon-crossing noise~\cite{Starobinsky1986,StarobinskyYokoyama1994} is
\emph{dry}~\cite{Morikawa2022}: it lacks a retarded partner generated by the effective
action, and therefore carries no genuine dissipation.

With self-interaction the two-loop CTP effective action generates both a retarded
kernel $\Sigma_R$ and the noise kernel (see Appendix~\ref{app:dS})
\begin{equation}
  N_\lambda(x,x')\sim \frac{\lambda^2}{6}\,a^3(t)a^3(t')
  \left[G_H^{\IR}(x,x')\right]^3 ,
  \label{eq:dSNoiseKernel}
\end{equation}
from the same sunset diagram.  The Langevin equation
\begin{equation}
  \left[\partial_t^2+3H\partial_t-a^{-2}\nabla^2\right]\phi
  +V'_{\eff}(\phi)
  +\int \dd^4x'\,\Sigma_R(x,x')\phi(x')
  = -\xi_\lambda(x)
  \label{eq:dSLangevin}
\end{equation}
is the non-dry stochastic dynamics.  In momentum space $N_\lambda\propto k^{-3}$ in
the infrared, reproducing the Harrison--Zel'dovich spectrum.  The key distinction from
standard stochastic inflation is that the spectrum is \emph{dynamical}---generated by
interaction---rather than kinematical.

The curvature perturbation amplitude~\cite{Mukhanov1992,LiddleLyth2000}
\begin{equation}
  \Delta_\zeta^2\simeq\frac{H^2}{8\pi^2\epsilon\mpl^2}
\end{equation}
is the cosmological analogue of the current noise coefficient $A_J$: both are the
observable coefficients of an infrared logarithm, and both are determined by the
microscopic parameters of the source.

Beyond the Gaussian leading order, the $\lambda\phi^4$ interaction generates
higher-order CTP vertices ($\phi_r\phi_a^3$ and above) that make the noise
\emph{multiplicative}---the noise amplitude depends on the field value $\phi_r$.  The
resulting non-Gaussian tails in the curvature perturbation distribution are discussed
in Section~\ref{sec:higher}, where their connection to primordial black hole formation
is noted.

\section{Higher loops, resummation, and non-Gaussian noise}
\label{sec:higher}

The rule for higher loops is simple: do not put an IRD back into a deterministic
action.  Construct the CTP effective action, separate its causal and symmetric parts,
represent the symmetric part stochastically.

At quadratic order in $q_a$ the result is Gaussian noise.  Higher loops produce higher
powers of $q_a$,
\begin{equation}
  \Gamma_{\rm ng}[q_r,q_a]
  = \sum_{n\geq 3}\frac{\ii^{\,n-1}}{n!}
  \int \dd x_1\cdots \dd x_n\,
  C_n(x_1,\ldots,x_n;q_r)\,
  q_a(x_1)\cdots q_a(x_n) ,
  \label{eq:nonGaussianCTP}
\end{equation}
corresponding to non-Gaussian stochastic cumulants $\langle
\xi(x_1)\cdots\xi(x_n)\rangle_c=C_n$.

In QED, leading soft divergences
exponentiate~\cite{YennieFrautschiSuura1961,Weinberg1965,Kibble1968,KulishFaddeev1970}.
Higher-loop soft divergences should be organized as corrections to the retarded
current-current kernel, the symmetric noise kernel \eqref{eq:QEDnoiseKernel}, possible
higher current cumulants, and finite matching coefficients.  This avoids double counting
of self-energy and vertex contributions.

In gravity, higher loops must respect the transversality constraint
\eqref{eq:transversality} at every order.  Near-critical phases and gravitational
collapse can produce strongly non-Gaussian metric fluctuations requiring the full
cumulant expansion.

\paragraph{Multiplicative noise and non-Gaussianity in de Sitter $\lambda\phi^4$ theory.}
The de Sitter case with $\lambda\phi^4$ interaction deserves special attention beyond
the leading two-loop result of~\cite{Morikawa2022}.  The quartic vertex couples four
powers of $\phi$, so in Keldysh variables the interaction generates terms of the form
$\phi_r^3\phi_a$ (classical drift) and $\phi_r\phi_a^3$ (beyond-Gaussian CTP term).
The latter produces a contribution to $\Gamma_{\rm ng}$ \eqref{eq:nonGaussianCTP} with
$n=4$.  At the level of the stochastic equation, this means that the noise strength
itself depends on the field value $\phi_r$---the noise is \emph{multiplicative}.

A multiplicative stochastic Langevin equation for the order parameter $\phi$,
\begin{equation}
  \dot\phi = -\frac{V'_{\eff}(\phi)}{3H}
  + \eta_0(t)
  + \eta_\lambda(t)\,g(\phi) ,
  \qquad
  \langle\eta_\lambda(t)\eta_\lambda(t')\rangle = N_\lambda\,\delta(t-t') ,
  \label{eq:multiplicativeLangevin}
\end{equation}
with $g(\phi)\propto\phi$, differs fundamentally from an additive Gaussian process.  The
probability distribution $P(\phi,t)$ satisfies a Fokker--Planck equation with
field-dependent diffusion coefficient, which generically produces
\emph{non-Gaussian tails}---specifically, the distribution decays more slowly than
Gaussian for large $|\phi|$, often approaching a power-law or exponential-tail form.

\paragraph{Connection to primordial black holes.}
This non-Gaussian tail has direct observational consequences.  The abundance of
primordial black holes (PBHs) is exponentially sensitive to the tail of the curvature
perturbation distribution $P(\mathcal{R})$: even a modest enhancement of the tail beyond
the Gaussian prediction can increase the PBH abundance by many orders of magnitude.
Recent work on stochastic inflation in the ultra-slow-roll
regime~\cite{Ezquiaga2020,Figueroa2021,Pattison2021} has demonstrated that
non-Gaussian exponential tails are a generic consequence of quantum diffusion in
non-linear potentials, and that they dramatically enhance PBH formation compared to
Gaussian estimates.

In the present CTP framework, the mechanism is transparent.  The $\lambda\phi^4$
interaction generates a noise kernel $N_\lambda(x,x')$ \eqref{eq:dSNoiseKernel} and,
through the $\phi_r\phi_a^3$ term, a non-vanishing fourth cumulant $C_4$.  At the
level of the probability distribution, $C_4\neq 0$ corresponds to excess kurtosis:
the distribution of $\phi$ has heavier tails than a Gaussian with the same variance.
Combined with the long-memory structure of the $1/f$-like infrared kernel, the rare
large-$\phi$ events are further enhanced compared to a Markovian Gaussian estimate.
The CTP effective-action framework therefore identifies a direct path from the
$\lambda\phi^4$ IRD to non-Gaussian tails that could source PBH
overproduction~\cite{Bullock1997,Hooshangi2022}.

The systematic computation of $C_3$ and $C_4$ from the two-loop and higher CTP
effective action for $\lambda\phi^4$ de Sitter scalar theory, and the resulting
prediction for the PBH mass spectrum, is a natural extension of the framework
developed here and of the results in~\cite{Morikawa2022}.

\section{Observational consequences and falsifiability}
\label{sec:observational}

The measurable quantity is not the divergence itself, but the coefficient of the induced
classical observable $S_{\rm obs}(\omega)=|\chi(\omega)|^2 N_{\rm IR}(\omega)$.
Table~\ref{tab:observables} summarizes the four arenas.

\begin{table}[htbp]
\centering
\small
\begin{tabularx}{\textwidth}{@{}YYYY@{}}
\toprule
System & IR sector & Noise kernel & Observable \\
\midrule
Soft QED conductor & soft photons
  & $N\propto ne^2\beta^2/|\omega|$
  & current $1/f$ noise; $A_J(n_{\rm imp})$ \\[4pt]
Soft gravitons & stress-tensor fluct.
  & $\mathcal{N}\propto GE^2/|\omega|$
  & $\Delta h^{\rm mem}$, $S_h(f)$ \\[4pt]
Phase transition & order-param.\ modes
  & $N_{\rm PT}\sim \lambda^2[G^H_{\IR}]^3$
  & $S(k)$, $\chi_{\rm ord}$, domains \\[4pt]
de Sitter inflation & long-$\lambda$ scalar
  & $N_\lambda\propto \lambda^2[G^{\IR}_H]^3$
  & $\Delta_\zeta^2(k)$, $n_s$, $C_\ell$ \\
\midrule
Many-body IR cloud & particle-hole pairs
  & Anderson-type $1/|\omega|$
  & $A(\omega)$, Loschmidt $G(t)$ \\
\bottomrule
\end{tabularx}
\caption{Infrared sectors and their observable classical signatures in the CTP
framework.  The many-body row is discussed in Appendix~\ref{app:Handel}.}
\label{tab:observables}
\end{table}

\paragraph{QED conductor (primary prediction).}
The key falsifiable prediction is the impurity-density dependence \eqref{eq:AJmain}.
In the dilute-impurity regime with $C_{\rm IR}\propto n_{\rm imp}^{s}$ and
$\gamma_{\rm imp}\propto n_{\rm imp}$,
\begin{equation}
  A_J(n_{\rm imp})\propto n_{\rm imp}^{s-2} .
\end{equation}
The minimal soft-dressing estimate ($s=1$, one soft-dressing event per scattering) gives
$A_J\propto n_{\rm imp}^{-1}$.
A direct experimental protocol would use a controlled family of samples with varying
impurity density while controlling carrier density, geometry, and temperature.  The
low-frequency noise should be fitted as
\begin{equation}
  S_I(f)=\frac{A_I}{f^\alpha}+S_I^{\rm white}+S_I^{\rm other}(f) ,
\end{equation}
with the infrared QED contribution expected to have $\alpha\simeq 1$.  Additional
checks include: (i)~a finite bandwidth over which $\alpha\simeq 1$; (ii)~systematic
dependence of $A_I$ on impurity density and mobility; (iii)~logarithmic bandwidth
dependence $\langle\delta I^2\rangle_{\rm IR}\simeq A_I\ln(f_{\max}/f_{\min})$; 
(iv)~independence from temperature at $T\to 0$ (distinguishing from Johnson--Nyquist
noise~\cite{Johnson1928,Nyquist1928});
and (v)~the absence of an intrinsic
low-frequency cutoff: the flattening point of the spectrum should track
$\omega_{\IR}\sim T_{\obs}^{-1}$ as the observation time is extended, without
saturating.  Observation of a $T_{\obs}$-independent flattening frequency
$\omega_c$ would falsify the soft-photon mechanism, whereas superpositions of
Lorentzians with a bounded relaxation-time distribution necessarily produce one.
The robustness of the $1/f$ signal under
propagation and nonlinear (thresholded) readout, relevant to its appearance across many
downstream systems, has been examined in the wave-packet
model~\cite{MorikawaNakamichi2023pink}.

\paragraph{Gravitational observables.}
The soft-graviton contribution \eqref{eq:gravNoise} implies metric noise
$S_h(f)\propto GE^2/f$.  It also predicts a permanent memory displacement
$\Delta h^{\rm mem}$, which is detectable in principle by pulsar timing arrays and
future space-based interferometers \cite{Christodoulou1991,Favata2010}.

\section{Relation to coarse graining, decoherence, and prior work}
\label{sec:discussion}

\paragraph{Coarse graining.}
The standard route~\cite{CalzettaHu2008,HuVerdaguer2008} chooses a system--environment
split and traces over the environment.  The self-selection principle of
Section~\ref{sec:ctp} makes this step a re-description rather than a necessity: in a
system with an instability or infrared divergence, the order parameter emerges
dynamically and the remaining modes act as its effective environment, so the split is
made by the dynamics itself.  The Hubbard--Stratonovich identity then represents the
noise algebraically, even when no environment has been designated by hand.

\paragraph{Decoherence.}
The imaginary influence action suppresses CTP histories with large $q_a$, favoring a
classical stochastic description.  No separate decoherence postulate is needed---but a
genuine system--environment (or self-selected order-parameter/effective-environment)
split is needed, and this is the operative distinction with respect to
Ref.~\cite{HsiangHu2022}, which shows that a closed, unitarily evolving quantum system
with no bipartition into system and environment cannot decohere intrinsically from
squeezing alone.  That result is correct and is not in tension with the present
mechanism: it is the statement that applies \emph{before} any order-parameter/
effective-environment split has occurred.  The self-selection principle of
Section~\ref{sec:ctp} is precisely what supplies such a split dynamically, without it
being imposed by hand; decoherence and the Langevin noise $N$ then arise together, from
the same imaginary part of $\Gamma$, once the split exists.  Neither can appear without
it, which is why Ref.~\cite{HsiangHu2022}, working in a single global Hilbert
space with no such split, correctly finds neither.  Whether the stochastic equation
becomes a macroscopic record, an order parameter, or a cosmological perturbation is a
subsequent physical question.  In the itinerant-vacuum language, decoherence is the
statement that different classical histories---corresponding to different vacuum
trajectories---cannot interfere: the orthogonality of the dressed states is the same
phenomenon as decoherence, viewed from the Hilbert-space rather than the density-matrix
perspective.  This identification is quantitatively the same one that underlies
horizon-induced decoherence~\cite{DanielsonSatishchandranWald2023} and infrared quantum
information~\cite{Carney2017,Carney2018,GomezLetschkaZell2018a,GomezLetschkaZell2018b},
where the soft which-path memory plays the
role of the environment record.

\paragraph{Two complementary perspectives on the infrared.}
It is useful to state plainly where the present work sits relative to the standard
particle-physics treatment of infrared divergences.  In the inclusive $S$-matrix
perspective, soft divergences cancel (Bloch--Nordsieck/KLN), dressed states restore
infrared finiteness, and the soft sector is a coherent cloud inseparable from the
charge; from this viewpoint there is no stochastic dynamics, and the no-go reasoning
of~\cite{Fukuyama2025} applies.  Fukuyama's later detector-resolution analysis
\cite{Fukuyama2026} is also useful here, because it emphasizes that observable
infrared memory is tied to finite resolution rather than to an uncancelled inclusive
cross section.  In the real-time, non-perturbative perspective adopted
here---and shared, in different languages, by stochastic
gravity~\cite{CalzettaHu2008,HuVerdaguer2008}, the infrared
triangle~\cite{Strominger2018,StromingerZhiboedov2016}, horizon
decoherence~\cite{DanielsonSatishchandranWald2023}, and infrared-finite scattering
theory~\cite{PrabhuSatishchandranWald2022}---the soft sector carries physical memory and
information, and an order parameter coupled to it acquires genuine stochastic dynamics.
The two perspectives are not in conflict; they answer different questions about the same
infrared structure.  The contribution of this paper is to make the real-time perspective
directly (through the Hubbard--Stratonovich identity), unified (across four arenas), and
predictive (through the $1/f$ current-noise amplitude).

\paragraph{Checks required in concrete applications.}
\begin{enumerate}[label=(\roman*)]
  \item Use the in--in CTP effective action, not the in--out action.
  \item Decompose the IRD-bearing kernel into a causal retarded part and a positive
    symmetric part before varying the action.
  \item Respect Ward identities \eqref{eq:Ward}, diffeomorphism transversality
    \eqref{eq:transversality}, and internal-symmetry constraints on the noise kernel.
  \item Verify the \emph{two} conditions for a genuine classical stochastic dynamics.
    (a)~Do not use dry noise: any stochastic source must have an interaction-generated
    retarded partner.  A conformally protected free field (e.g.\ source-free Maxwell in
    de Sitter) is dry and does not classicalize, in agreement with~\cite{Fukuyama2025};
    the interacting sector coupled to a charged current is not.
    (b)~Verify anticommutator dominance: the classical statistical reading of the
    Hubbard--Stratonovich field is faithful only where the occupation ratio
    \eqref{eq:occupationRatio} is large.  Infrared-divergent sectors satisfy this
    automatically ($n\to\infty$ in the soft limit); a finite vacuum Hadamard kernel
    ($n\simeq0$) is quantum noise, not classical statistics.
  \item Confirm that the observable is a real-time order-parameter correlator, not an
    inclusive cross section: BN/KLN cancellation removes the infrared part of the
    latter but not of the former (Section~\ref{sec:fukuyama}).
  \item For $1/f$ behavior, $\omega_{\IR}\sim T_{\obs}^{-1}$; $\omega_{\max}$ is the
    shortest physical time in the infrared scaling regime.
\end{enumerate}

\section{Conclusion}
\label{sec:conclusion}

Infrared divergences are not merely pathologies of perturbation theory.  They are
signatures of an itinerant vacuum: a quantum vacuum that wanders continuously through
a family of inequivalent states as the classical order parameter evolves under the
stochastic force generated by the CTP effective action.  When properly relocated from
the deterministic part of the action into the stochastic part via the
Hubbard--Stratonovich identity, an IRD becomes the origin of observable classical
structure.  The relocation is an algebraic identity; no prior coarse graining
or decoherence is required.

The four arenas of this paper illustrate different facets of the mechanism.
\begin{description}
\item[Soft QED and $1/f$ current noise.]
The soft-photon phase space generates a $1/f$ noise kernel for conserved currents,
Eq.~\eqref{eq:QEDnoiseKernel}, with amplitude $A_J\propto ne^2\beta^2/\gamma_{\rm
imp}^2$---a constrained prediction in terms of carrier density, charge, velocity, and
the impurity scattering rate.
This is the field-theoretic foundation of Handel's intuition~\cite{Handel1975,Handel1980},
now formulated as a CTP identity that resolves the formal objections
of~\cite{Nieuwenhuizen1987,KissHeszler1986,Weissman1988}: soft-sector orthogonality is
reduced-state decoherence, not a contradiction.  The recent no-go
result~\cite{Fukuyama2025} is fully respected: it concerns inclusive cross sections and
the conformally protected free field, whereas the present noise lives in the real-time
correlator of an interacting current---a complementary object that carries memory and
is not constrained by BN/KLN cancellation (Section~\ref{sec:fukuyama}).

\item[Soft gravitons and gravitational memory.]
The stress-tensor Hadamard kernel produces $1/f$ metric noise with amplitude
$\propto GE^2/|\omega|$.  Gravitational memory is the retarded kernel at $\omega\to 0$.
Memory and metric noise are inseparable faces of the same infrared singularity.

\item[Non-equilibrium phase transitions.]
The infrared instability of the order parameter generates a multiplicative colored-noise
Langevin equation.  Macroscopic classicality arises from the CTP separation, not from
externally imposed decoherence.

\item[De Sitter scalar fields and primordial black holes.]
The two-loop $\lambda\phi^4$ noise kernel reproduces the Harrison--Zel'dovich
spectrum~\cite{Morikawa2022}.  The free-field kinematical noise is dry; the
interaction-generated noise is the genuine dynamical source.  Beyond the Gaussian
leading order, the quartic vertex generates multiplicative noise through the
$\phi_r\phi_a^3$ CTP vertex, producing non-Gaussian tails in the probability
distribution of the curvature perturbation.  These tails could source primordial black
hole overproduction~\cite{Ezquiaga2020,Figueroa2021,Bullock1997}, connecting the CTP
infrared mechanism directly to one of the most actively studied topics in current
cosmology.
\end{description}

Higher-loop IRDs follow the same rule: organize into retarded kernels and Gaussian or
non-Gaussian noise cumulants.  The broader message, summarized in
\begin{equation}
  \text{infrared divergence}
  \;\longrightarrow\;
  \text{itinerant vacuum}
  \;\longrightarrow\;
  \text{classical stochastic degrees of freedom} ,
\end{equation}
connects QED soft photons, de Sitter fluctuations, gravitational memory, and critical
phenomena within a common CTP framework.

In every case, the infrared divergence is not merely a defect of the theory but a sign
that the vacuum is changing its state.  The CTP effective action reads this sign
correctly: it separates the divergence into a retarded memory of past transitions and a
stochastic force toward future ones.  The divergence is thereby
converted from an embarrassment into an instrument of prediction.

The two ingredients of this construction---a Langevin equation for the mean field read
off directly from the CTP effective action, with non-Gaussian and colored noise, and
the separation of an infrared-divergent kernel by the identity alone---were both
present, the first explicitly and the second in embryo, in the author's earliest work on
dissipative quantum systems~\cite{Morikawa1986,Morikawa1987}.  The present paper returns
to that origin and carries it to its natural conclusion: the infrared divergence is not
a defect to be removed but the seed of an itinerant vacuum.

\paragraph{Outlook: geometry of the vacuum manifold.}
The itinerant-vacuum picture suggests a natural geometric extension that we sketch
here and develop elsewhere.  As the order parameter $\phi(t)$ traces a stochastic
trajectory in the full configuration space $\mathcal{M}_\phi$, a distinct Hilbert
space $\mathcal{H}[\phi_0]$ is attached to \emph{every} point
$\phi_0\in\mathcal{M}_\phi$ as a fibre---not only to the stable minima.  Each fibre
is defined by the deterministic part of the Lagrangian evaluated at $\phi_0$; the
stochastic force is then a fibre-changing perturbation.  The full structure is a
Hilbert-space bundle
\begin{equation}
  \bigsqcup_{\phi_0\in\mathcal{M}_\phi}\mathcal{H}[\phi_0]
  \;\longrightarrow\; \mathcal{M}_\phi ,
  \label{eq:HilbertBundle}
\end{equation}
in which the dressed-state overlap $D_{ij}$ plays the role of a connection, encoding
parallel transport between neighbouring fibres.

For a Mexican-hat potential $U(\phi)=\lambda(|\phi|^2-v^2)^2$ with complex $\phi$,
the base space is the entire complex $\phi$-plane,
$\mathcal{M}_\phi\cong\mathbb{C}\cong\mathbb{R}^2$.  The familiar vacuum manifold
$\mathrm{VM}\cong S^1$ of radius $|\phi|=v$ is a special submanifold of
$\mathcal{M}_\phi$: the locus of absolute minima.  But the origin ($\phi=0$, the
unstable symmetric point) and any point on the potential hill between origin and $S^1$
are equally valid base points, each with its own healthy Hilbert space $\mathcal{H}[\phi_0]$.
The Goldstone infrared divergence is the flatness of the bundle connection along $S^1$;
the retarded kernel encodes parallel transport between fibres; and the noise kernel
drives the stochastic geodesic from one fibre to another.

This fibre-bundle picture resolves several otherwise awkward problems.
\emph{Unstable particles}, whose states formally escape a fixed Hilbert space
(necessitating Gamow vectors or Rigged Hilbert spaces), are naturally described as
trajectories on $\mathcal{M}_\phi$ that drift away from an unstable fixed point
toward a distant fibre---without invoking a non-standard inner product.
\emph{Quantum measurement} can be formulated as a stochastic process on the bundle:
einselection~\cite{Zurek2003} localises the trajectory onto a preferred fibre (a
pointer state), while the procoherence mechanism of~\cite{MorikawaNakamichi2006}
drives the order parameter toward a definite broken-symmetry vacuum on $S^1$, with
the classical meter reading identified as the angular coordinate $\theta$ on the
vacuum manifold.  The CTP Langevin equation derived in this paper is then the
equation of motion for the meter pointer itself.

A systematic development of this geometric framework---including the differential
geometry of $\mathcal{M}_\phi$, the curvature of the Hilbert-space bundle, and its
applications to measurement theory and the classification of infrared
sectors---is left to a forthcoming paper.

A separate cosmological question is whether the stress-energy fluctuations of ordinary
matter, acting as a source through the gravitational sector of
Section~\ref{sec:gravity}, leave an imprint on the dark sector of the late universe.
Because such an imprint would track the matter that sources it, it could bear on the
coincidence problem; but establishing its equation of state and magnitude requires a
dedicated quantitative study and lies beyond the scope of this paper.

\appendix
\section{Hubbard--Stratonovich identity}
\label{app:HS}

For a real positive kernel $N$, the Gaussian identity is
\begin{equation}
  \int \mathcal{D}\xi\,
  \exp\!\left[-\frac{1}{2}\xi N^{-1}\xi + \ii q_a\xi\right]
  = \mathcal{N}'\exp\!\left[-\frac{1}{2}q_aNq_a\right] .
\end{equation}
Completing the square,
\begin{equation}
  -\frac{1}{2}(\xi-\ii Nq_a)N^{-1}(\xi-\ii Nq_a)
  -\frac{1}{2}q_aNq_a .
\end{equation}
If higher powers of $q_a$ are present, the stochastic process is non-Gaussian and is
specified by its cumulant functional.

\section{Useful formulae for the \texorpdfstring{$1/f$}{1/f} kernel}
\label{app:oneoverf}

With the convention \eqref{eq:oneoverfSpectrum},
\begin{equation}
  N_{1/f}(t)=\frac{A}{\pi}\left[\operatorname{Ci}(\omega_{\max}|t|)
  -\operatorname{Ci}(\omega_{\IR}|t|)\right] .
\end{equation}
For $\omega_{\IR}|t|\ll1$,
$\operatorname{Ci}(\omega_{\IR}|t|)=\gamma_E+\ln(\omega_{\IR}|t|)+\order((\omega_{\IR}t)^2)$.
For $\omega_{\max}|t|\gg1$, $\operatorname{Ci}(\omega_{\max}|t|)$ is oscillatory and
suppressed as $1/(\omega_{\max}|t|)$.  The memory is therefore logarithmic in the
intermediate range $t_{\UV}\ll t\ll T_{\obs}$.

\section{Soft QED as a Wilson-line influence functional}
\label{app:wilson}

For a charged particle with \emph{eikonal current}
$j^\mu(x)=e\int \dd\tau\,u^\mu(\tau)\delta^{(4)}(x-z(\tau))$---i.e., the current sourced
by a fixed, undeflected classical worldline $z(\tau)$ with four-velocity $u^\mu(\tau)$,
the approximation in which the recoil of the emitting charge under photon emission is
neglected---the gauge-invariant form of \eqref{eq:QEDInfluence} is
\begin{equation}
  \Gamma_\gamma = -\ln\!\left\langle
  \mathcal{P}\exp\!\left[\ii e\int_{C_+}\dd z^\mu A_\mu\right]
  \mathcal{P}\exp\!\left[-\ii e\int_{C_-}\dd z^\mu A_\mu\right]
  \right\rangle_A .
  \label{eq:WilsonLine}
\end{equation}
Under a gauge transformation $A_\mu\to A_\mu+\partial_\mu\lambda$, each path-ordered
exponential acquires only a boundary phase $\exp[\ii e(\lambda(z_f)-\lambda(z_i))]$; since
the $C_+$ and $C_-$ contours share the same endpoints $z_i,z_f$, these phases cancel
between the two exponentials for any $\lambda$, at every order in $e$.  Equation
\eqref{eq:WilsonLine} is therefore gauge invariant non-perturbatively, without appeal to
\eqref{eq:Ward} order by order: expanding in powers of $e$ generates every attachment of
a photon line to the worldline, with no distinction between what would be called a
``self-energy'' or a ``vertex'' insertion in the usual Feynman-diagram language: all such
insertions are attachment points on the same line.  Equation~\eqref{eq:Ward} is the
statement, valid order by order, that the sum of self-energy and vertex cuts is
gauge invariant; \eqref{eq:WilsonLine} is the non-perturbative, all-orders object whose
gauge invariance implies \eqref{eq:Ward} rather than needing to be checked against it.
Cutting \eqref{eq:WilsonLine} on shell---replacing the exchanged photon line(s) by their
Wightman functions, as in the general discussion of Section~\ref{sec:ctp}---therefore
gives the gauge-invariant $D^H_{\mu\nu}$ of \eqref{eq:HadamardPhoton} directly, with no
separate self-energy/vertex bookkeeping.

Expanding \eqref{eq:WilsonLine} to lowest order in $e$ attaches a single soft photon of
momentum $k$ to the worldline; the near-on-shell charged propagator pole combines with
the vertex to give the \emph{eikonal factor}
\begin{equation}
  e\,\frac{u^\mu}{u\cdot k} ,
  \label{eq:eikonalFactor}
\end{equation}
universal (spin- and process-independent) because it depends only on the charge and
velocity of the emitting line, not on its internal structure---the field-theoretic
statement of Low's theorem~\cite{Low1958}.  It is this factor, squared and summed over sources, that
produces the denominators $(u_i\cdot k)(u_j\cdot k)$ below; the angular dependence on
$x=\cos\theta$ in $f(\beta)$ is a consequence of \eqref{eq:eikonalFactor}, not its
definition.  For $N=nV$ isotropic sources with speed $\beta$, the current-projected
noise kernel in the soft limit is
\begin{equation}
  N^{\mu\nu}_{\rm QED}(k)
  = \sum_{i,j} e^2\,
  2\pi\delta(k^2)\,
  \frac{u_i^\mu u_j^\nu - u_i\cdot u_j\, P^T_{\mu\nu}}%
   {(u_i\cdot k/|k|)(u_j\cdot k/|k|)} .
\end{equation}
After the angular integral and particle sum,
\begin{equation}
  N^{\mu\nu}_{\rm QED}(\omega)
  = n\,e^2\,
  \frac{f(\beta)}{|\omega|}\,P^T_{\mu\nu} ,
\end{equation}
with the velocity factor arising from the transverse-projected eikonal integral
\begin{equation}
  f(\beta)=\frac{1}{2}\int_{-1}^{1}\!\dd x\,
  \frac{\beta^2(1-x^2)}{(1-\beta x)^2}
  =\frac{1}{\beta}\ln\frac{1+\beta}{1-\beta}-2
  \xrightarrow{\beta\to0}\frac{2}{3}\beta^2 ,
\end{equation}
which is \eqref{eq:velFactor}.  The $1/|\omega|$ follows from the radial measure
$\omega\,\dd\omega$ times the eikonal $1/(p\cdot k)^2\propto1/\omega^2$.  This is
Eq.~\eqref{eq:QEDnoiseKernel}.

\paragraph{The dressing as a current-driven coherent state.}
It is worth noting why the Kulish--Faddeev dressing~\cite{KulishFaddeev1970,Kibble1968}
appears automatically in this picture.  A classical c-number current $j^\mu$ coupled
linearly to the free photon field drives each field mode like a forced oscillator, whose
ground state is a Glauber coherent state with displacement $\alpha_k\sim
\tilde\jmath(k)/\sqrt{2\omega_k}$, i.e.\ an infinite superposition of soft photons.  For
the eikonal current $j^\mu=e\!\int\!\dd\tau\,u^\mu\delta^{(4)}(x-z)$ this displacement is
$\alpha_k\sim e\,u\cdot\epsilon/(u\cdot k)$, identical to the Kulish--Faddeev soft factor.
The dressed asymptotic state, introduced in the $S$-matrix formulation as a prescription
to restore infrared finiteness, is therefore nothing but the coherent state that a
classical charged current necessarily produces.  The same background-current (Wilson
line) model has been analyzed by Hirai and Sugishita~\cite{HiraiSugishita2021}, who
showed that it shares the full infrared structure of QED and used it to demonstrate
that the original Kulish--Faddeev dressing alone does not render the $S$-matrix
infrared finite: the correct, gauge-invariant dress code is fixed by the asymptotic
symmetry~\cite{HiraiSugishita2019,HiraiSugishita2023}.  That refinement concerns the
infrared finiteness of $S$-matrix elements, which the present construction does not
invoke; the noise-kernel derivation above uses the coherent state only as the field
configuration driven by a classical current, and is insensitive to the choice among
infrared-equivalent dressings.  Its mean photon number $\langle
N_\gamma\rangle\sim e^2\!\int\!\dd\omega\,f(\beta)/\omega$ is the same infrared logarithm
that appears in the noise kernel above.  In the itinerant-vacuum language, the order
parameter (the current) labels a coherent vacuum; as it evolves it drives a succession
of such vacua, and averaging over the unresolved soft history of this succession yields
the stochastic kernel.  The single dressing of a fixed scattering event and the noise of
an evolving order parameter are the same coherent-state physics viewed statically and
dynamically.

\section{Soft gravitons and stress-tensor noise}
\label{app:graviton}

The gravitational kernel follows the same steps as Appendix~\ref{app:wilson} with the
replacements spin-1 $\to$ spin-2, charge $\to$ energy-momentum, and $P^T_{\mu\nu}\to
\Lambda_{\mu\nu\alpha\beta}$.  The soft-graviton factor carries two powers of momentum,
$\kappa\,p^\mu p^\nu/(p\cdot k)$ with $\kappa^2=32\pi G$, because the source is the
stress tensor $T^{\mu\nu}=\sum_i m_i\!\int\!\dd\tau_i\,u_i^\mu u_i^\nu\,
\delta^{(4)}(x-z_i)$ rather than the current.  The transverse-traceless projector is the
spin-2 analogue of $P^T_{\mu\nu}$,
\begin{equation}
  \Lambda_{\mu\nu\alpha\beta}
  = \tfrac{1}{2}\!\left(P^T_{\mu\alpha}P^T_{\nu\beta}
  + P^T_{\mu\beta}P^T_{\nu\alpha} - P^T_{\mu\nu}P^T_{\alpha\beta}\right) .
  \label{eq:TTprojector}
\end{equation}
The same on-shell phase space and diagonal-source sum give
\begin{equation}
  \mathcal{N}^{\rm soft}_{\mu\nu\alpha\beta}(\omega)
  = n_g\,(32\pi G)\,\frac{E^2\,f_g(\beta)}{|\omega|}\,
  \Lambda_{\mu\nu\alpha\beta}(\hat n) ,
  \label{eq:gravNoiseFull}
\end{equation}
the full form of \eqref{eq:gravNoise}, with energy $E=\gamma m$ replacing the charge and
the spin-2 velocity factor
\begin{equation}
  f_g(\beta)=\frac{1}{2}\int_{-1}^{1}\!\dd x\,
  \frac{\beta^4(1-x^2)^2}{(1-\beta x)^2}
  \xrightarrow{\beta\to0}\frac{8}{15}\beta^4 .
  \label{eq:fgrav}
\end{equation}
The quadratic $\beta^2$ of QED (dipole radiation) is replaced by the quartic $\beta^4$
(quadrupole radiation), the only structural change between the two arenas.  Static mass
($\beta\to0$) does not radiate: only the time-varying quadrupole contributes to the
noise, consistent with the absence of monopole and dipole gravitational radiation.

\paragraph{Temporal versus spatial spectrum: one infrared divergence.}
The $1/|\omega|$ spectra of QED and gravity, and the $1/k^3$ Harrison--Zel'dovich
spectrum of the de~Sitter field (Section~\ref{sec:ds}), are two faces of the same
infrared divergence, related by the on-shell condition $\omega=|\bm k|$.  A noise kernel
$\propto\delta(k^2)/(p\cdot k)^2$ gives, in the temporal description, the power spectral
density $N(\omega)\propto1/|\omega|$; in the equal-time spatial description, the
three-dimensional power spectrum
\begin{equation}
  P(k)\sim\frac{1}{k}\cdot\frac{1}{k^2}=\frac{1}{k^3},
\end{equation}
where the $1/k$ is the on-shell phase space and the $1/k^2$ is the soft (eikonal)
factor.  The two statements are equivalent expressions of the logarithmic infrared
divergence,
\begin{equation}
  \int\frac{\dd\omega}{\omega}
  \;\Longleftrightarrow\;
  \int\frac{k^2\,\dd k}{k^3}=\int\frac{\dd k}{k} .
\end{equation}
A flat temporal power ($\omega N=\text{const}$) and a scale-invariant spatial power
($k^3 P=\text{const}$) are the same statement.  Thus gravitational metric noise is $1/f$
when observed as a time series (pulsar timing, interferometry) and Harrison--Zel'dovich,
$P(k)\propto1/k^3$, when observed as a spatial pattern---the same itinerant-vacuum
infrared structure as the de~Sitter spectrum, reached here through soft gravitons rather
than super-horizon scalar modes.

\section{Scaling estimate for the soft-photon noise coefficient \texorpdfstring{$C_{\rm IR}$}{CIR}}
\label{app:CIR}

For a charged-particle scattering event with momentum transfer $\Delta p$, the soft
emission probability is
\begin{equation}
  \dd P_{\rm soft}=\mathcal{C}_{\rm soft}\,\alpha_{\rm em}\,\frac{\dd\omega}{\omega} ,
\end{equation}
where $\mathcal{C}_{\rm soft}$ contains angular and kinematic factors.  If scattering
events occur at rate $\Gamma_{\rm scatt}$, the infrared stochastic force strength scales as
\begin{equation}
  C_{\rm IR}\sim \alpha_{\rm em}\,\mathcal{C}_{\rm soft}\,\Gamma_{\rm scatt}\,(\Delta p)^2 .
  \label{eq:CIRscaling}
\end{equation}
For dilute impurities, $\Gamma_{\rm scatt}\simeq n_{\rm imp}v_F\sigma_{\rm tr}\simeq
\gamma_{\rm imp}$.  Substituting into \eqref{eq:AJmain},
\begin{equation}
  A_J\sim \alpha_{\rm em}\,\mathcal{C}_{\rm soft}\,\left(\frac{ne}{m}\right)^2
  \frac{(\Delta p)^2}{\gamma_{\rm imp}} .
\end{equation}
If $\Delta p$ and $\sigma_{\rm tr}$ are independent of impurity density, this gives
$A_J\propto n_{\rm imp}^{-1}$.  This is a scaling estimate.  A complete derivation
must compute $N_{\rm IR}^{\mu\nu}$ in a screened, finite-temperature, disordered
electron gas.

\section{Relation to Handel's infrared theory of \texorpdfstring{$1/f$}{1/f} noise}
\label{app:Handel}

Handel proposed that $1/f$ noise in electric currents is a fundamental QED phenomenon
associated with soft-photon radiative corrections to charged-particle
scattering~\cite{Handel1975,Handel1980}.  This was an important and suggestive idea: it
identified a possible connection between the universality of low-frequency current noise
and the universal soft structure of QED.

\paragraph{Objections to Handel's theory.}
The proposal was criticized by several authors~\cite{Nieuwenhuizen1987,KissHeszler1986,Weissman1988}.
The central objections are:
\begin{enumerate}
  \item Scattering states with different soft-photon content are orthogonal in
    the infrared limit (soft-sector orthogonality):
    \begin{equation}
      {}_{\rm soft}\langle\gamma_j|\gamma_i\rangle_{\rm soft}
      \to 0 \quad \text{as } \omega_{\IR}\to 0 .
    \end{equation}
    This makes it problematic to identify interference between different soft sectors as
    an ordinary current fluctuation in an $S$-matrix framework.
  \item A direct identification of infrared-divergent cross sections with current
    fluctuations is not rigorously justified.
\end{enumerate}

\paragraph{The objections as constructive contributions.}
The criticisms of Refs.~\cite{Nieuwenhuizen1987,KissHeszler1986,Weissman1988} deserve
appreciation as physically precise analyses that sharpen the question rather than close
it.  Nieuwenhuizen et al.\ and Kiss--Heszler correctly identified that soft-sector
orthogonality makes interference between different dressing states ill-defined in an
$S$-matrix framework: this is not a mere technical complaint, but a genuine physical
observation about the structure of QED Hilbert space.  Weissman's comprehensive
review~\cite{Weissman1988} further established that any proposal connecting QED
infrared structure to $1/f$ noise must provide a rigorous bridge between cross sections
and fluctuation spectra---a bridge that Handel's original formulation lacked.

These are not objections to be dismissed; they are structural constraints that the
correct theory must satisfy.  The present CTP formulation is designed to satisfy all
of them.  Together, Handel's intuition and its critics collectively charted the
logical territory of the problem.

\paragraph{The CTP resolution.}
In the present formulation both objections are resolved.  For a physical charged state
dressed by soft photons,
\begin{equation}
  |\Psi\rangle = \sum_i c_i\,|p_i\rangle_{\rm hard}\otimes|\gamma_i\rangle_{\rm soft} ,
\end{equation}
tracing over unresolved photons gives the reduced density matrix with decoherence
factors
\begin{equation}
  \rho_{\rm hard}^{ij}=c_ic_j^*\,
  D_{ij}\,|p_i\rangle\langle p_j| ,
  \qquad
  D_{ij}={}_{\rm soft}\langle\gamma_j|\gamma_i\rangle_{\rm soft}
  =\exp\!\left[-\frac{1}{2}B_{ij}\ln\frac{\omega_{\max}}{\omega_{\IR}}
  +\ii\Phi_{ij}\right] .
  \label{eq:Dij}
\end{equation}
When the soft clouds become orthogonal, $D_{ij}\to 0$ and interference between
different hard histories is suppressed.  This is reduced-state \emph{decoherence}, not
a failure of the description.  At any finite resolution the suppression is partial:
$D_{ij}$ remains nonzero, with a magnitude controlled by the kinematics and by the
infrared resolution scale, as analyzed in detail
in~\cite{GomezLetschkaZell2018a,GomezLetschkaZell2018b}; complete orthogonality holds
only in the strict limit $\omega_{\IR}\to0$.  Objection (1) is therefore not a
contradiction but a
statement that the soft sector carries which-path information.  In the itinerant-vacuum
language of this paper, the orthogonality $D_{ij}\to 0$ simply means that a charged
particle that has moved from momentum $p_i$ to $p_j$ now inhabits a genuinely
\emph{different vacuum}: the dressed coherent state has shifted from $|\gamma[p_i]\rangle$
to $|\gamma[p_j]\rangle$, and these are orthogonal because the vacuum has moved.  
Note that the exponent $B_{ij}\ln(\omega_{\max}/\omega_{\IR})$ in \eqref{eq:Dij} is the same
$\int\dd\omega/\omega$ integral that makes $\langle N_\gamma\rangle$ diverge: the
divergence of the soft occupation and the orthogonality of the dressed vacua are the
same infrared fact, viewed as a number and as an overlap. 
The
$1/f$ noise is the classical trace of this continuous vacuum wandering.

Objection (2) is resolved by replacing the direct cross-section argument with the CTP
chain \eqref{eq:chain}.  The reduced in-in dynamics is written as a Schwinger--Keldysh
influence action.  The infrared part of its imaginary component is
\begin{equation}
  \mathrm{Im}\,\Gamma_{\rm IF}^{\rm IR}
  = \frac{1}{2} J_\Delta\, N_{\rm IR}\, J_\Delta .
\end{equation}
By the Hubbard--Stratonovich identity,
\begin{equation}
  \exp\!\left[-\mathrm{Im}\,\Gamma_{\rm IF}^{\rm IR}\right]
  = \int \mathcal{D}\xi\,\mathcal{P}[\xi]\,
  \exp\!\left(\ii\int \xi J_\Delta\right) ,
  \qquad
  \langle\xi(\omega)\xi(-\omega)\rangle=N_{\rm IR}(\omega) .
\end{equation}
The route to $1/f$ noise is \eqref{eq:chain}, not a direct identification of an
infrared-divergent cross section with a current fluctuation.

\paragraph{Physical analogy: Anderson orthogonality catastrophe.}
The soft-photon decoherence factor \eqref{eq:Dij} is structurally identical to the
Anderson orthogonality catastrophe in a Fermi sea~\cite{Anderson1967,Mahan2000},
\begin{equation}
  |\langle\Psi_0'|\Psi_0\rangle|^2\sim
  \exp\!\left[-\alpha_{\rm AOC}\ln\frac{E_{\UV}}{E_{\IR}}\right] ,
\end{equation}
where $\alpha_{\rm AOC}$ is set by scattering phase shifts and
the infrared cloud is made of low-energy particle-hole pairs.  Observable consequences
include Fermi-edge singularities $A(\omega)\sim(\omega-\omega_{\rm th})^{-\alpha_{\rm FES}}$
and Loschmidt echoes~\cite{NozieresDeDominicis1969,Hentschel2005}.  In a disordered
conductor, both soft photons and particle-hole pairs may contribute to low-frequency
noise:
\begin{equation}
  N_{\rm total} = N_{\rm soft\ photon} + N_{\rm particle\text{-}hole}
  + N_{\rm defects} + \cdots .
\end{equation}
The present paper focuses on the soft-photon contribution; the particle-hole term is an
important complementary channel that deserves a parallel CTP treatment.

\paragraph{Scaling comparison with Handel.}
At the level of scaling, the present approach preserves Handel's central intuition.  The
coefficient $A_J$ is controlled by charged-particle scattering events that update the
soft electromagnetic dressing.  The difference is interpretation: Handel attributed the
noise to a divergent cross section, whereas here it arises from the trace over the
soft sector encoded in the CTP influence action.  The fine-structure-constant dependence
$A_J\propto\alpha_{\rm em}$ (via $C_{\rm IR}\propto\alpha_{\rm em}$, see
Appendix~\ref{app:CIR}) is present in both pictures, but the amplitude formula
\eqref{eq:AJmain} also contains the impurity-density and velocity factors absent from
Handel's universal coefficient.

\section{Interaction-generated de Sitter noise kernel}
\label{app:dS}

For $\lambda\phi^4$ theory in de Sitter space, the CTP contour interaction is
\begin{equation}
  S_{\rm int}[\phi_+]-S_{\rm int}[\phi_-]
  = -\int \dd^4x\,a^3(t)\frac{\lambda}{4!}(\phi_+^4-\phi_-^4) .
\end{equation}
In Keldysh variables, $\phi_+^4-\phi_-^4=4\phi_r^3\phi_a+\phi_r\phi_a^3$.  The first
term gives the classical drift $V'(\phi_r)$; loop corrections generate retarded and
noise kernels.  The schematic two-loop structure
\begin{equation}
  N_\lambda(x,x')\propto \lambda^2 [G_H(x,x')]^3
\end{equation}
shows that noise requires interaction: a purely kinematical, dry stochastic source
cannot produce this kernel.

\vspace{6pt}
\section*{Acknowledgements}
The author is grateful to Akika Nakamichi for valuable discussions and for collaboration on earlier studies that laid the foundation for the present work.

\bibliographystyle{unsrtnat}
\bibliography{refs_v27}

\end{document}